\pdfoutput=1
\documentclass[pdflatex,sn-mathphys-num]{sn-jnl}

\usepackage{graphicx}
\usepackage{amsmath,amssymb,amsfonts}
\usepackage{amsthm}
\usepackage{mathrsfs}
\usepackage[title]{appendix}
\usepackage{mathtools}
\usepackage{physics}
\usepackage{xcolor}
\usepackage{url}
\usepackage{tikz}
\usepackage{orcidlink}
\usepackage{etoolbox}
\usetikzlibrary{positioning}
\IfFileExists{dsfont.sty}{\usepackage{dsfont}}{}

\definecolor{ao}{rgb}{0.0, 0.5, 0.0}

\numberwithin{equation}{section}
\renewcommand{\thesubsection}{\thesection(\alph{subsection})}

\makeatletter
\patchcmd{\@maketitle}{Corresponding author(s). E-mail(s): }{E-mail: }{}{\PackageWarning{main}{maketitle patch failed}}
\patchcmd{\email}{;\ }{}{}{\PackageWarning{main}{email patch failed}}
\makeatother
\begin{document}

\title[Bondi--Sachs formalism on past light cones]{Bondi--Sachs formalism on past light cones and an analogue of the integral mass formula}

\author*{\fnm{Matheus G.} \sur{Barbosa}\,\orcidlink{0000-0003-0595-9897}}\email{matheusgb@ifsc.usp.br}

\affil{\orgdiv{S\~{a}o Carlos Institute of Physics}, \orgname{University of S\~{a}o Paulo}, \orgaddress{\city{S\~{a}o Carlos}, \postcode{13566-590}, \state{SP}, \country{Brazil}}}

\abstract{Following a proposal that the thermodynamic aspects of general relativity can be found in the evolution of the past light cones of arbitrary worldlines, we generalise the Bondi--Sachs formalism to these surfaces. The observer may accelerate and rotate their tetrad, the spacetime may contain matter and a cosmological constant, and no symmetry is assumed. We impose regularity conditions at the vertex of each light cone and along the polar axis, including general axial conditions that allow for shear. Given a suitable prescription for the stresses of matter, the Einstein field equations can then be integrated hierarchically from data on a single light cone, and the regularity at the vertex fixes all constants of integration. The solution reveals a universal property: to first order in the observer's proper time, the evolution of the metric and of the four-momentum crossing the light cones is independent of the specific behaviour of matter. As an application, we derive, for an observer immersed in matter, an analogue of the integral mass formula of black hole mechanics.}


\maketitle
{\renewcommand{\thefootnote}{}\footnotetext{This article presents part of the results of the author's doctoral thesis, submitted to the examination committee at the S\~{a}o Carlos Institute of Physics, University of S\~{a}o Paulo. A preprint containing the complete set of results is available as arXiv:2609.25477 \cite{Barbosa2026}.}}

\section{Introduction}\label{intro}

This work is part of a broader programme on the thermodynamic aspects of general relativity \cite{Barbosa2026}. Black hole thermodynamics \cite{Bardeen1973,Bekenstein1973,Hawking1975} and its extensions to cosmological horizons \cite{Gibbons1977} and to the Rindler horizon \cite{Unruh1976} suggest that these aspects are associated with causal horizons in general. Inspired by previous proposals based on local Rindler horizons \cite{Jacobson1995,Jacobson2003,Padmanabhan2010a}, the author's thesis~\cite{Barbosa2026} suggested that these aspects can be found in the evolution of the past light cones associated with an arbitrary worldline, which that work regards as the basic geometric structure underlying the more usual causal horizons. Since the observer is represented by a single worldline and the relevant thermodynamic system is the content of their past light cone at each moment, the description is naturally adapted to the notion of `photographic' thermodynamics proposed in~\cite{Dunkel2009}.

In this work, we generalise the Bondi--Sachs formalism \cite{Bondi1962,Sachs1962,VanderBurg1966} to these past light cones and use it to derive an analogue of the integral mass formula of black hole mechanics \cite{Bardeen1973}. The construction is general: the worldline may be accelerated, the observer's orthonormal tetrad may rotate relative to Fermi--Walker transport, the spacetime may contain matter and a cosmological constant, and no symmetry is assumed. The asymptotic conditions of the original formulation, imposed at null infinity, are replaced by regularity conditions at the vertex of each light cone, which we call central conditions, and along the polar axis of the angular coordinates. All quantities are thus defined relative to the observer, which ties the formalism directly to what they can measure.

This work extends the results of~\cite{Maedler2013}, which derived the regularity conditions of the Bondi--Sachs metric at the vertex of a null cone traced by a timelike geodesic, for axially symmetric vacuum spacetimes. The non-axisymmetric case was later considered for the linearised vacuum equations \cite{Maedler2013b}. Our construction treats the full nonlinear equations with matter and a cosmological constant, for a worldline that need not be a geodesic and a tetrad that may rotate, so that the leading terms of the metric functions $U$, $W$, and $\Phi$ at the vertex no longer vanish and central conditions are also needed for the energy-momentum tensor. We focus on two aspects: the first orders of the metric, which are fixed by the regularity conditions and by the kinematics of the worldline, and the axial conditions in spacetimes without axial symmetry, where shear along the axis makes them nontrivial. The higher orders, including the curvature-dependent coefficients obtained in~\cite{Maedler2013} for axially symmetric vacuum spacetimes, are left to the hierarchical scheme that solves the Einstein field equations.

Since the first law of black hole thermodynamics is connected with the Komar mass \cite{Bardeen1973,Komar1959}, which can be interpreted as the source of the gravitational field, we analyse the leading-order term of the source of the gravitational potential, identified in the Newtonian limit and extended to the general case. This yields the active gravitational mass seen by the observer, read off from an expansion of the metric at the observer, much as the original Bondi mass is read off near null infinity \cite{Bondi1962}. Together with a proposed scalar surface element for past light cones, the formalism leads to an equation with the same structure as the integral mass formula for nonrotating black holes, a first point of contact with the thermodynamic aspects of general relativity. This equation can also be regarded as an analogue, for expansions in positive powers of $r$ around the vertex, of the mass of future light cones of~\cite{Chrusciel2014}, which is read off from expansions in negative powers of $r$ near null infinity.

Our main results are the following. The central conditions fix all constants of integration of the hierarchical scheme: the only nonvanishing ones carry the gauge freedom of the worldline and its tetrad, and none plays the role of the news function. General axial conditions, which allow for shear along the axis, replace the trivial one used under axial symmetry. As a small improvement of the proof of~\cite{Isaacson1983}, we show explicitly that ten equations of the hierarchical scheme recover all the Einstein field equations. To first order in the observer's proper time, the evolution of the metric and of the four-momentum crossing the light cones is independent of the specific behaviour of matter, so that different interactions of matter enter only at second order. Finally, the analogue of the integral mass formula has an area term identical in form to that of a nonrotating black hole, including the numerical factor, and a fictitious contribution fixed by the observer's acceleration and rotation.

The paper is organised as follows. Section~\ref{coorsys} constructs the coordinate system and section~\ref{regcond} derives the regularity conditions at the vertex and along the axis. Section~\ref{EFEs} presents the hierarchical integration of the field equations, the constants of integration, and spherically symmetric spacetimes as an example. Section~\ref{imf} derives the analogue of the integral mass formula, and section~\ref{conclusions} summarises the results. Appendices~\ref{dustapp} and~\ref{kinematics} treat spherically symmetric dusty spacetimes and the kinematics of light cones, and Appendices~\ref{appfs} and~\ref{firstterms} contain the complete field equations and the first terms of the solution.

The software \textit{Wolfram Mathematica} was used to compute many of the results in this work. The suite \textit{xAct} \cite{xAct}, in particular, the packages \textit{xCoba} and \textit{xTensor}, was used to compute curvatures, Christoffel symbols, and the field equations. Curvature calculations were checked with a code kindly provided by D. A. T. Vanzella (personal communication, February 10, 2023). Figures~\ref{z} and \ref{shear} were drawn with \textit{Inkscape}. AI models (Claude Opus 5.5 and Claude Fable 5.2, Anthropic) were used to prepare this article from the author's thesis: to select and condense the material, to draft the abstract, the introduction, and the conclusions, to reformulate the derivation of the axial conditions in section~\ref{axial}, to edit the language, and to suggest references. They also verified all equations independently of the \textit{Mathematica} computations, with \textit{Python} scripts that evaluate them numerically for generic metrics, compare series expansions, and rederive the remaining results. All derivations and results are the author's, who reviewed the whole text and takes full responsibility for it.

Throughout this work, the term \emph{light cone} refers to past light cones. We use Planck units, $ c = G = \hbar = k_B = 1 $, and the metric signature $(-,+,+,+)$. Abstract indices are lowercase Latin letters $a,\ldots,h$, and components carry Greek indices; $\mu,\nu,\ldots\in \{u,r,\theta,\phi\}$ in Bondi--Sachs coordinates, $i,j,k\in\{1,2,3\}$ label non-timelike components, and $A, B \in \{\theta,\phi\}$ label angular components. The summation convention applies only to pairs of contravariant-covariant indices.

\section{The coordinate system}\label{coorsys}

Let $\mathcal{P}\colon \mathbb{R} \supseteq (a,b) \to \mathcal{M}$ be a $C^3$, future-directed timelike curve in the spacetime $\mathcal{M}$, contained in a convex normal neighbourhood $ \mathcal{N} $ \cite{Hawking1973,Chrusciel2011}, parametrised by proper time $ \tau $, with four-velocity $v^a$, four-acceleration $a^a$, and an angular velocity $\Omega^a$ relative to Fermi--Walker transport \cite[p. 174]{Misner1973}, both orthogonal to $v^a$. The most general rule for propagating an orthonormal basis $ \{ e_{(\bar{\mu})}{}^a \} $ along $ \mathcal{P} $, with $ e_{(\bar{0})}{}^a = v^a $, is Fermi--Walker transport plus a rotation \cite{Padmanabhan2010b}:
\begin{equation}\label{tetradtransport}
v^a \nabla_a e_{(\bar{\mu})}{}^b = e_{(\bar{\mu})}{}_a (v^b a^a - v^a a^b +  v_c \Omega_d \epsilon^{cdab}),
\end{equation}
where $ \epsilon_{abcd} $ is the Levi-Civita tensor, with $ \epsilon_{0123} = -\epsilon^{0123} = + 1 $ \cite[p. 87]{Misner1973}.

For each unit spacelike vector $n^a = n^{\bar{i}} e_{(\bar{i})}{}^a$ at $\mathcal{P}(\tau)$, we define angles $\theta'$ and $\phi'$ by
\begin{equation}\label{n}
n^{\bar{\mu}} \eqqcolon \left(0,\sin(\theta')\cos(\phi'),\sin(\theta')\sin(\phi'),\cos(\theta') \right),
\end{equation}
and keep the components $n^{\bar{i}}$ constant along $\mathcal{P}$. The past-directed null vector
\begin{equation}\label{k}
\left .{k^a(\theta',\phi')}\right|_{\mathcal{P}} \coloneqq \left .{n^a(\theta',\phi')-v^a}\right|_{\mathcal{P}}
\end{equation}
generates a null geodesic, affinely parametrised by $\lambda'$ with $\lambda' = 0$ at $\mathcal{P}(\tau)$. The geodesics in all directions form the past light cone $\mathcal{C}(\tau)$ of $\mathcal{P}(\tau)$, and the function $u'$ with $u'|_{\mathcal{C}(\tau)} = \tau$ completes a coordinate system $ y^{\mu'} = (u', \lambda', \theta', \phi') $ on a neighbourhood $ \mathcal{N}' \subseteq\mathcal{N} $ of $ \mathcal{P} $. Since $\partial_{(\lambda')}{}^a = k^a$ is null, $g_{\lambda'\lambda'} = 0$. Moreover, $k_a\nabla_b k^a = 0$, $[\partial_{(A')},k]^a = 0$, and the geodesic equation imply $k^b \nabla_b (k_a \partial_{(A')}{}^a) = 0$, so $k_a \partial_{(A')}{}^a$ is constant along each generator. To find its value, consider the observer's proper coordinates $x^{\bar{\mu}}$ \cite[Sec.~13.6]{Misner1973}, whose coordinate basis coincides with $\{e_{(\bar{\mu})}{}^a\}$ at $\mathcal{P}(\tau)$, and assume that the metric components are $C^3$ functions of them. Then, $x^{\bar{\mu}} = x^{\bar{\mu}}|_{\mathcal{P}(\tau)} + k^{\bar{\mu}}|_{\mathcal{P}(\tau)} \lambda' + \mathcal{O}({\lambda'}^2)$, so that $\partial_{(A')}{}^{\bar{\mu}} = \partial_{A'} n^{\bar{\mu}} \lambda' + \mathcal{O}({\lambda'}^2)$ and, since $k_{\bar{\mu}}|_{\mathcal{P}(\tau)}\partial_{A'} n^{\bar{\mu}} = n_{\bar{i}}\partial_{A'} n^{\bar{i}} = 0$, $k_a \partial_{(A')}{}^a = \mathcal{O}({\lambda'}^2)$. Being constant along the generator, it must vanish; hence $g_{\lambda' A'} = 0$, and each light cone is a null hypersurface \cite{Gourgoulhon2006}.

\begin{figure}
	\centering
	\begin{tikzpicture}
	\node[anchor=south west,inner sep=0] (image) at (0,0) {\includegraphics[width=0.48\textwidth]{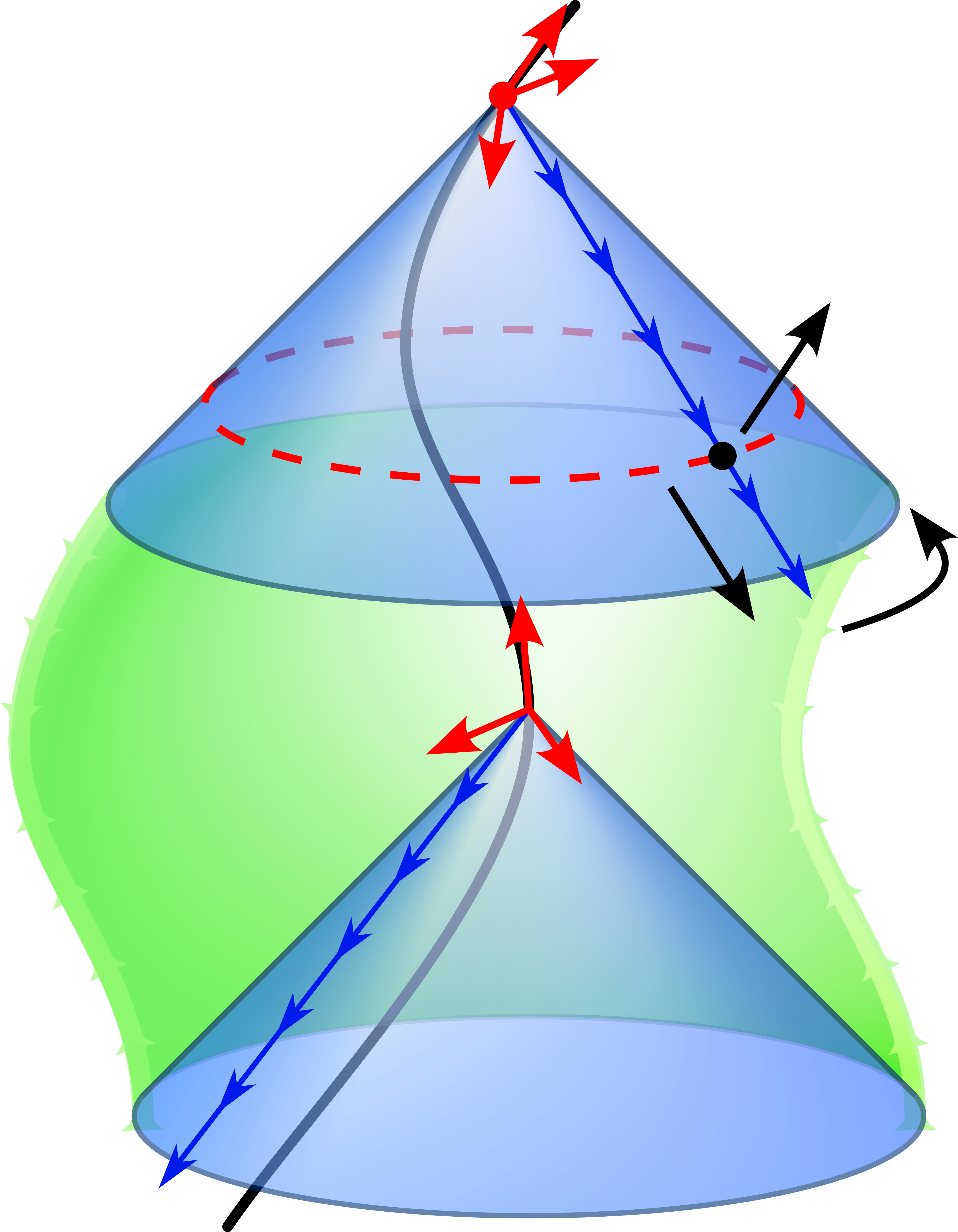}};
	\begin{scope}[x={(image.south east)},y={(image.north west)}]
	\node[anchor=west] at (0.240,-0.003) {$\mathcal{P}$};
	\node[anchor=north,text=blue] at (0.123,0.042) {$k^a(\theta,\phi)$};
	\node[anchor=south east,text=red] at (0.510,0.391) {$e_{(\bar{1})}{}^a$};
	\node[anchor=south west,text=red] at (0.592,0.357) {$e_{(\bar{2})}{}^a$};
	\node[anchor=west,text=red] at (0.557,0.481) {$e_{(\bar{0})}{}^a$};
	\node[anchor=west,text=ao] at (0.041,0.397) {$\simeq \mathbb{R} \times S^2$};
	\node[anchor=north west] at (0.768,0.492) {$r$};
	\node[anchor=west] at (0.990,0.537) {$\phi$};
	\node[anchor=south west] at (0.850,0.746) {$u$};
	\node[anchor=north west] at (0.779,0.644) {$\mathcal{Q}$};
	\node[anchor=east,text=red] at (0.205,0.678) {$\mathcal{S}(u,r)$};
	\node[anchor=north,text=red] at (0.522,0.859) {$e_{(\bar{1})}{}^a$};
	\node[anchor=north west,text=red] at (0.580,0.943) {$e_{(\bar{2})}{}^a$};
	\node[anchor=east,text=red] at (0.592,0.999) {$e_{(\bar{0})}{}^a$};
	\node[anchor=east,text=red] at (0.522,0.926) {$\mathcal{P}(\tau)$};
	\node[anchor=east] at (0.346,0.999) {$\boxed{\theta = \text{constant}}$};
	\end{scope}
	\end{tikzpicture}
	\caption[Bondi--Sachs coordinates]{Bondi--Sachs coordinates. The coordinates $ z^\mu $ are constructed around the worldline $ \mathcal{P} $, along which the observer propagates an orthonormal basis $ \left\{ e_{(\bar{\mu})}{}^a \right\} $. Each past light cone is generated by null geodesics, such as those in blue, which have $k^a(\theta,\phi)$ as the tangent vector field and constant values of $ \theta $ and $ \phi $. At each light cone of constant $ u $, we define a foliation by surfaces $ \mathcal{S}(u,r) \simeq S^2 $ so that their areas equal $ 4\pi r^2 $. Then, a point $ \mathcal{Q} $ on the past light cone of $ \mathcal{P}(\tau) $ will be identified by the coordinates $ z^\mu = (u,r,\theta,\phi) $, with $ u=\tau $. In this picture, $ \theta $ is held constant while the other coordinates increase in the indicated directions. Moreover, since our analysis may be limited by the neighbourhood $ \mathcal{N} $ to finite values of $ r $, we shall consider only a region of spacetime bounded by some timelike hypersurface $\simeq \mathbb{R} \times S^2$, such as the one represented in green.}
	\label{z}
\end{figure}

The Bondi--Sachs coordinates $ z^\mu = (u,r,\theta,\phi) $ (figure~\ref{z}) are defined by $ u \coloneqq u'$, $\theta \coloneqq \theta'$, $\phi \coloneqq \phi'$, and the areal radius
\begin{equation}\label{r}
r \coloneqq \left(\frac{\det(g_{A'B'}(y^{\rho'}))}{\sin[2](\theta')}\right)^{1/4}
\end{equation}
so that $\det(g_{AB}) = r^4 \sin^2(\theta)$, while $g_{rr} = g_{rA} = 0$.
By defining
\begin{equation}\label{h}
h_{AB} \coloneqq r^{-2}g_{AB}, \qq{which has} \det(h_{AB}) = \sin[2](\theta),
\end{equation}
we write the metric, as in~\cite{Maedler2013,VanderBurg1966}, in terms of six functions $\Phi$, $\beta$, $U$, $W$, $\gamma$, and $\delta$ of all coordinates, with $U^A$ defined by $U^\theta \coloneqq U$ and $U^\phi \coloneqq W$. Since $k_a \propto \nabla_a u$, one has $g^{uu} = g^{uA} = 0$, and the metric and its inverse read
\begin{gather}\label{g}
\begin{matrix}
g_{\mu\nu} = \mqty(-e^{2(\Phi + \beta)} + r^2 h_{AB}U^A U^B & e^{2\beta} & - r^2 h_{CB}U^C \\ e^{2\beta} & 0 & 0\\ - r^2 h_{CA}U^C & 0 & r^2 h_{AB} ), \\[4ex]
h_{AB} = \mqty(e^{2\gamma}\cosh(2\delta) & \sinh(2\delta)\sin(\theta) \\ \sinh(2\delta)\sin(\theta) & e^{-2\gamma}\cosh(2\delta)\sin[2](\theta)) , \\[4ex]
g^{\mu\nu} = \mqty(0 & e^{-2\beta} & 0 \\ e^{-2\beta} & e^{2(\Phi - \beta)} & e^{-2\beta}U^B\\ 0 & e^{-2\beta}U^A & r^{-2} h^{AB}) , \\[4ex]
h^{AB} = \mqty(e^{-2\gamma}\cosh(2\delta) & -\sinh(2\delta)\csc(\theta) \\ -\sinh(2\delta)\csc(\theta) &  e^{2\gamma}\cosh(2\delta)\csc[2](\theta)) , 
\end{matrix}
\end{gather}
\begin{equation}\label{detg}
g \coloneqq \det(g_{\mu\nu})=-e^{4\beta}r^4 \sin[2](\theta).
\end{equation}

\section{Regularity conditions}\label{regcond}

\subsection{Central conditions on the metric}

The motion of the observer's tetrad determines the leading terms of the metric along $\mathcal{P}$. In the observer's proper reference frame, with proper coordinates $x^{\bar{\mu}} = (\tau, s n^{\bar{i}})$ built from spacelike geodesics of proper length $s$ orthogonal to $v^a$, equation \eqref{tetradtransport} fixes the Christoffel symbols along $\mathcal{P}$, and the metric reads \cite{Misner1973,Ni1978}
\[
d s^2 = -(1+2 a_{\bar{i}} x^{\bar{i}})\, {d x^{\bar{0}}}^2 +2 \epsilon_{\bar{i} \bar{j} \bar{k}} \Omega^{\bar{j}} x^{\bar{k}}\, d x^{\bar{0}} d x^{\bar{i}} +\delta_{\bar{i} \bar{j}}\, d x^{\bar{i}} d x^{\bar{j}}+ \mathcal{O}(x^{\bar{i}} x^{\bar{j}})\, d x^{\bar{\mu}} d x^{\bar{\nu}},
\]
where $a_{\bar{i}}$ and $\Omega^{\bar{j}}$ are evaluated at $x^{\bar{0}}$, and $\epsilon_{\bar{i} \bar{j} \bar{k}}$ is the three-dimensional Levi-Civita symbol. Solving the geodesic equation for the null generators with these Christoffel symbols, for a metric of class $C^3$ in proper coordinates, one finds
\begin{equation}\label{x0}
x^{\bar{0}}\left(u^{\prime}, \lambda^{\prime}, \theta^{\prime}, \phi^{\prime}\right)=u^{\prime}-\lambda^{\prime}+a_{\bar{k}}\left(u^{\prime}\right) n^{\bar{k}}\left(\theta^{\prime}, \phi^{\prime}\right) \lambda^{\prime 2}+\mathcal{O}\left(\lambda^{\prime 3}\right)
\end{equation}
and
\begin{equation}\label{xi}
x^{\bar{i}}\left(u^{\prime}, \lambda^{\prime}, \theta^{\prime}, \phi^{\prime}\right)=n^{\bar{i}}\left(\theta^{\prime}, \phi^{\prime}\right) \lambda^{\prime}+\left[\epsilon_{\bar{i} \bar{j} \bar{k}} \Omega^{\bar{j}}\left(u^{\prime}\right) n^{\bar{k}}\left(\theta^{\prime}, \phi^{\prime}\right)-\frac{a^{\bar{i}}\left(u^{\prime}\right)}{2}\right] \lambda^{\prime 2}+\mathcal{O}\left(\lambda^{\prime 3}\right).
\end{equation}
We transform the metric to the coordinates $y^{\mu'}$ with the derivatives of these expansions, omitting the curvature terms, which appear at second order in proper coordinates \cite{Manasse1963,Ni1978} and will be determined by the Einstein field equations. Since the angular coordinates are dimensionless, each angular index raises the order in $\lambda'$ by one, and
$$
\begin{aligned}
g_{ u^{\prime} u^{\prime}}&=-\left[1+2 a_{\bar{i}} n^{\bar{i}} \lambda^{\prime}\right]+\mathcal{O}	\left({\lambda'}^2\right), \\
g_{u^{\prime} \lambda^{\prime}}&=1+\mathcal{O}\left({\lambda'}^2\right), \\
g_{u^{\prime} A^{\prime}}&=\left(\epsilon_{\bar{i}\bar{j}\bar{k}} \partial_{A'} n^{\bar{i}} \Omega^{ \bar{j}} n^{\bar{k}}-a_{\bar{k}} \partial_{A'} n^{\bar{k}}\right) {\lambda'}^2+\mathcal{O}\left({\lambda'}^3\right), \\
g_{A^{\prime} B^{\prime}}&={\lambda'}^2 \partial_{A'} n^{\bar{i}} \partial_{B'} n_{\bar{i}}+\mathcal{O}\left({\lambda'}^4\right),
\end{aligned}
$$
with $a_{\bar{i}}$ and $\Omega^{\bar{i}}$ evaluated at $\tau = u'$, since $a_{\bar{i}}(x^{\bar{0}}) = a_{\bar{i}}(u') + \mathcal{O}(\lambda')$ and $\Omega^{\bar{i}}(x^{\bar{0}}) = \Omega^{\bar{i}}(u') + \mathcal{O}(\lambda')$ for a $C^3$ curve. The angular part is, at leading order, the metric of a sphere of radius $\lambda'$, so $\det(g_{A'B'}) = {\lambda'}^4\sin^2(\theta') + \mathcal{O}({\lambda'}^6)$, and equation \eqref{r} gives
\begin{equation}\label{ytoz}
u=u^{\prime}, \quad r=\lambda^{\prime}+\mathcal{O}\left({\lambda'}^3\right),\quad \theta=\theta^{\prime},\quad \text {and} \quad \phi=\phi^{\prime} .
\end{equation}
With the Jacobian of this transformation, the metric in Bondi--Sachs coordinates has the same form,
$$
\begin{aligned}
g_{ u u}&=-\left[1+2 a_{\bar{i}} n^{\bar{i}} r\right]+\mathcal{O}	\left(r^2\right), \\
g_{u r}&=1+\mathcal{O}\left(r^2\right), \\
g_{uA}&=\left(\epsilon_{\bar{i}\bar{j}\bar{k}} \partial_{A} n^{\bar{i}} \Omega^{ \bar{j}} n^{\bar{k}}-a_{\bar{k}} \partial_{A} n^{\bar{k}}\right) r^2+\mathcal{O}\left(r^3\right), \\
g_{A B}&=r^2 \partial_{A} n^{\bar{i}} \partial_{B} n_{\bar{i}}+\mathcal{O}\left(r^4\right).
\end{aligned}
$$
We now assume that the metric functions can be expanded around $r = 0$ up to some order $N$ as\footnote{Differentiability up to order $N$ implies that the remainder is $o(r^N)$ \cite[p. 287]{Apostol1967}, and hence $\mathcal{O}(r^N)$, the more usual notation, which we use throughout. We assume, without stating it each time, that all functions are differentiable to the order needed for the expansions and derivatives used.}
\begin{equation}\label{f}
f\left(z^\mu\right)=\sum_{\xi=0}^N f^{(\xi)}\left(u, z^A\right) \frac{r^{\xi}}{\xi !}+\mathcal{O}\left(r^{N}\right) ,
\end{equation}
and compare the components of the metric \eqref{g} with the expansions above, which gives
\begin{gather*}
\gamma^{(0)}=\gamma^{(1)}=\delta^{(0)}=\delta^{(1)}=0, \\
U^{(0)}=a_{\bar{k}} \partial_{\theta} n^{\bar{k}} -\epsilon_{\bar{i}\bar{j}\bar{k}} \partial_{\theta} n^{\bar{i}} \Omega^{ \bar{j}} n^{\bar{k}}, \\
\sin^2 (\theta) W^{(0)}=a_{\bar{k}} \partial_{\phi} n^{\bar{k}} -\epsilon_{\bar{i}\bar{j}\bar{k}} \partial_{\phi} n^{\bar{i}} \Omega^{ \bar{j}} n^{\bar{k}}, \\
\beta^{(0)}=\beta^{(1)}=\Phi^{(0)}=0 \quad\qand\quad \Phi^{(1)}=a_{\bar{i}} n^{\bar{i}} .
\end{gather*}
Writing all sums explicitly, we obtain the \emph{central conditions} on the metric:
\begin{equation}\label{ccg}
\begin{gathered}
\gamma=\mathcal{O}\left(r^2\right),\quad \delta=\mathcal{O}\left(r^2\right), \quad \beta=\mathcal{O}\left(r^2\right), \\
U=\cos( \theta)\left(\cos (\phi) a_{\bar{1}}+\sin (\phi) a_{\bar{2}}\right)-\sin (\theta) a_{\bar{3}}+\sin (\phi) \Omega^{\bar{1}} -\cos(\phi)\Omega^{\bar{2}} + \mathcal{O}(r),\\
W =\csc (\theta)\left(-\sin(\phi)  a_{\bar{1}}+\cos(\phi)a_{\bar{2}}\right)+\cot (\theta)\left(\cos( \phi) \Omega^{\bar{1}}+\sin(\phi)\Omega^{\bar{2}}\right)-\Omega^{\bar{3}} +\mathcal{O}(r),\\
\Phi=\left[\sin (\theta)\left(\cos (\phi) a_{\bar{1}}+\sin( \phi) a_{\bar{2}}\right)+\cos( \theta) a_{\bar{3}}\right]  r+\mathcal{O}\left(r^2\right) \text {, }
\end{gathered}
\end{equation}
with the terms proportional to the Riemann tensor \cite{Maedler2013} still undetermined. As we will see, $ U^{(0)} $ and $ W^{(0)} $ are constants of integration. It can also be shown that $ \Phi^{(1)} $ is enforced by the Einstein field equations. For a geodesic, nonrotating observer, $U^{(0)}$, $W^{(0)}$, and $\Phi^{(1)}$ vanish, as in~\cite{Maedler2013}.

\subsection{Central conditions on the energy-momentum tensor}

We assume that the components of the energy-momentum tensor can be written as Taylor polynomials around $x^{\bar{i}}=0$ in proper coordinates:
\begin{equation}\label{expT}
T^{\bar{\mu}\bar{\nu}}(x^{\bar{\rho}}) = T^{\bar{\mu}\bar{\nu}}(x^{\bar{0}},0,0,0) + \mathcal{O}(x^{\bar{i}}).
\end{equation}
To transform this expansion, note that the angles $\theta'$ and $\phi'$ are propagated along the null generators, while the proper coordinates use spacelike geodesics. Let $\bar{\theta}$ and $\bar{\phi}$ be the extensions of the angles at $\mathcal{P}$ that are constant along each spacelike geodesic, so that $x^{\bar{i}} = n^{\bar{i}}(\bar{\theta},\bar{\phi})\, s$, with $s = (\delta_{\bar{i}\bar{j}}x^{\bar{i}}x^{\bar{j}})^{1/2}$. Then
\begin{equation}\label{thetaphi}
\left\{{\theta'},{\phi'}\right\}  =  \left.\left\{\bar{\theta},\bar{\phi}\right\}\right|_{\mathcal{P}} + \mathcal{O}(s), 
\end{equation}
and equation \eqref{xi} gives $\lambda' = s + \mathcal{O}(s^2)$. Since $s$, $\bar{\theta}$, and $\bar{\phi}$ do not depend on $x^{\bar{0}}$, these relations and equation \eqref{x0} imply
\begin{equation}\label{d0}
\partial_{\bar{0}} \lambda' = \mathcal{O}({\lambda'}^2), \qquad \partial_{\bar{0}} y^{A'} = \mathcal{O}(\lambda'), \qquad \partial_{\bar{0}} u' = 1 + \mathcal{O}({\lambda'}^2).
\end{equation}
For the spatial derivatives, $\partial_{\bar{i}} s = x^{\bar{i}}/s$, so that
\begin{equation}\label{di}
\partial_{\bar{i}} \lambda' = n^{\bar{i}}(\theta',\phi') + \mathcal{O}(\lambda') \qquad\text{and}\qquad \partial_{\bar{i}} u' = n^{\bar{i}}(\theta',\phi') + \mathcal{O}(\lambda'),
\end{equation}
where the second relation follows from equation \eqref{x0}. Writing $\bar{\theta} = \arccos(x^{\bar{3}}/s)$ and $\bar{\phi} = \arctan(x^{\bar{2}}/x^{\bar{1}})$, one also finds
\begin{equation}\label{dithetaphi}
\begin{aligned}
\partial_{\bar{i}} \theta' &= \frac{\cot(\theta')n^{\bar{i}}(\theta',\phi') - \csc(\theta')\delta^{\bar{3}}{}_{\bar{i}}}{\lambda'} + \mathcal{O}(1),          \\
\partial_{\bar{i}} \phi' &= \frac{\csc(\theta')(\cos(\phi')\delta^{\bar{2}}{}_{\bar{i}} - \sin(\phi')\delta^{\bar{1}}{}_{\bar{i}})}{\lambda'} + \mathcal{O}(1),
\end{aligned}
\end{equation}
away from the polar axis. By equation \eqref{ytoz}, the derivatives of the Bondi--Sachs coordinates with respect to the proper coordinates are
\begin{equation}\label{dbarbondi}
\begin{gathered}
\partial_{\bar{0}} u = 1 + \mathcal{O}(r^2),  \qquad \partial_{\bar{0}} r = \mathcal{O}(r^2), \qquad \partial_{\bar{0}} z^A = \mathcal{O}(r), \\
\partial_{\bar{i}} u = n^{\bar{i}}(\theta,\phi) + \mathcal{O}(r), \qquad \partial_{\bar{i}} r = n^{\bar{i}}(\theta,\phi) + \mathcal{O}(r),\\
\partial_{\bar{1}} \theta = \frac{\cos(\theta)\cos(\phi)}{r} + \mathcal{O}(1), \qquad \partial_{\bar{2}} \theta = \frac{\cos(\theta)\sin(\phi)}{r} + \mathcal{O}(1), \\
\partial_{\bar{3}} \theta = - \frac{\sin(\theta)}{r} + \mathcal{O}(1), \qquad \partial_{\bar{1}} \phi = - \frac{\csc(\theta)\sin(\phi)}{r} + \mathcal{O}(1),  \\
\partial_{\bar{2}} \phi = \frac{\csc(\theta)\cos(\phi)}{r} + \mathcal{O}(1) \quad\qand\quad \partial_{\bar{3}} \phi = \mathcal{O}(1). 
\end{gathered}
\end{equation}
Transforming $T^{\bar{\mu}\bar{\nu}}$ with these derivatives, each contravariant angular index contributes a factor $\mathcal{O}(1/r)$, and lowering it with the metric \eqref{g} contributes $\mathcal{O}(r^2)$. Hence, each covariant angular index raises the order by one, and
\begin{equation}\label{centralT}
T_{uA}{}^{(0)}=T_{rA}{}^{(0)}=T_{AB}{}^{(0)}=T_{AB}{}^{(1)}=0 \text {. }
\end{equation}
For later use, we give the components
\begin{equation}\label{expTBondi}
\begin{aligned}
T_{ru} &= -{T}^{\bar{0}\bar{0}{(0)}} - {T}^{\bar{0}\bar{i}{(0)}} n_{\bar{i}} + \mathcal{O}(r), \\
T_{rr} &= {T}^{\bar{0}\bar{0}{(0)}} + 2{T}^{\bar{0}\bar{i}{(0)}} n_{\bar{i}} + {T}^{\bar{i}\bar{j}{(0)}} n_{\bar{i}} n_{\bar{j}} + \mathcal{O}(r), \\
T_{rA} &= \left({T}^{\bar{0}\bar{i}{(0)}} + {T}^{\bar{i}\bar{j}{(0)}} n_{\bar{j}}\right) \partial_A n_{\bar{i}}\, r + \mathcal{O}(r^2),
\end{aligned}
\end{equation}
where $T^{\bar{\mu}\bar{\nu}(0)} = T^{\bar{\mu}\bar{\nu}(0)}(u) \coloneqq T^{\bar{\mu}\bar{\nu}}(x^{\bar{0}},0,0,0)$ and $n_{\bar{i}} = n_{\bar{i}}(\theta,\phi)$. Some of these components are not single-valued at the vertex \cite{Penrose1980}, because the direction of maximum increase in $r$ degenerates there, so that their values depend on the direction, specified by the angular coordinates, onto which the tensor is projected.

\subsection{Axial conditions}\label{axial}

A similar degeneracy occurs on the polar axis, where $\sin(\theta) = 0$: $\partial_{(\theta)}{}^a$ keeps a dependence on $\phi$ as $\sin(\theta)\to 0$, while $\partial_{(\phi)}{}^a \to 0^a$. Regularity conditions are needed to keep invariant quantities well defined there.

We assume that the map that sends each direction $n^{\bar{i}}$ at the vertex to the point where its generator meets $\mathcal{S}(u,r)$ is a $C^1$ diffeomorphism from $S^2$ onto this surface. Since $x \coloneqq \sin(\theta)\cos(\phi) = n^{\bar{1}}$ and $y \coloneqq \sin(\theta)\sin(\phi) = n^{\bar{2}}$ are smooth coordinates on $S^2$ near the poles, they are $C^1$ coordinates on $\mathcal{S}(u,r)$ near the axis, and we use $z^{\hat{\mu}} \coloneqq (u,r,x,y)$ as coordinates there. The two coordinate systems are related by
\begin{equation}\label{rel}
\begin{gathered}
\nabla_a\theta=\frac{\cos(\phi)\nabla_a x+\sin(\phi)\nabla_a y}{\cos(\theta)},\qquad
\nabla_a\phi=\frac{-\sin(\phi)\nabla_a x+\cos(\phi)\nabla_a y}{\sin(\theta)},\\
\partial_{(\theta)}{}^a=\cos(\theta)\left[\cos(\phi)\partial_{(x)}{}^a+\sin(\phi)\partial_{(y)}{}^a\right],\qquad
\frac{\partial_{(\phi)}{}^a}{\sin(\theta)}=-\sin(\phi)\partial_{(x)}{}^a+\cos(\phi)\partial_{(y)}{}^a .
\end{gathered}
\end{equation}

The metric functions can then be written in terms of the components of the metric in $z^{\hat{\mu}}$. By the inverse metric \eqref{g}, $e^{-2\beta}=g^{ur}$ and $e^{2(\Phi-\beta)}=g^{rr}$, so that $\beta$ and $\Phi$ are finite and single-valued on the axis. Similarly, $U^A=g^{rA}/g^{ru}$, so that the components $U^x=g^{rx}/g^{ru}$ and $U^y=g^{ry}/g^{ru}$ of $U^a \coloneqq U^A\partial_{(A)}{}^a$ are also finite and single-valued there. Relations \eqref{rel} give
\begin{equation}\label{Uw}
U=\frac{\cos(\phi)\,U^x+\sin(\phi)\,U^y}{\cos(\theta)},\qquad
w \coloneqq \sin(\theta)\,W=-\sin(\phi)\,U^x+\cos(\phi)\,U^y
\end{equation}
and, contracting the metric with $\partial_{(\theta)}{}^a$ and $\partial_{(\phi)}{}^a$,
\begin{equation}\label{gd}
\begin{aligned}
e^{2\gamma}\cosh(2\delta)&=\frac{\cos^2(\theta)}{r^2}\left[g_{xx}\cos^2(\phi)+g_{xy}\sin(2\phi)+g_{yy}\sin^2(\phi)\right],\\
\sinh(2\delta)&=\frac{\cos(\theta)}{r^2}\left[\tfrac12\left(g_{yy}-g_{xx}\right)\sin(2\phi)+g_{xy}\cos(2\phi)\right].
\end{aligned}
\end{equation}
Hence, $\gamma$, $\delta$, $U$, and $w$ are finite on the axis, whereas $W$ diverges as $1/\sin(\theta)$.

On the axis, $h_{xx} \coloneqq r^{-2}g_{xx}$, $h_{xy} \coloneqq r^{-2}g_{xy}$, and $h_{yy} \coloneqq r^{-2}g_{yy}$ are finite and single-valued, and, by relations \eqref{rel}, the condition $\det(h_{AB})=\sin^2(\theta)$ in \eqref{h} implies that $h_{xx}h_{yy}-h_{xy}^2=1$ there. Being symmetric, positive definite, and of unit determinant, the matrix of these components can be written as $R(\phi_0)\operatorname{diag}(e^{2\zeta},e^{-2\zeta})R(\phi_0)^{\mathsf{T}}$, where $R(\phi_0)$ is the rotation by an angle $\phi_0$, and $\zeta$ and $\phi_0$ depend only on $u$ and $r$. Equations \eqref{gd} then give, on the axis,
\begin{equation}\label{ax}
\begin{aligned}
\left.e^{2\gamma}\cosh(2\delta)\right|_{\mathcal{Q}_0}&=\cosh(2\zeta)+\sinh(2\zeta)\cos\!\left(2(\phi-\phi_0)\right),\\
\left.\sinh(2\delta)\right|_{\mathcal{Q}_0}&=\mp\sinh(2\zeta)\sin\!\left(2(\phi-\phi_0)\right),
\end{aligned}
\end{equation}
with the upper (lower) sign for $\cos(\theta)|_{\mathcal{Q}_0}=+1$ $(-1)$. These equations, together with equations \eqref{Uw} on the axis, will be called the \emph{general axial conditions}. They give the exact dependence of $\gamma$, $\delta$, $U$, and $w$ on $\phi$ along the axis, which corresponds to only four degrees of freedom: $\zeta$, $\phi_0$, $U^x|_{\mathcal{Q}_0}$, and $U^y|_{\mathcal{Q}_0}$.

\begin{figure}
	\centering
	\includegraphics[width=0.34\textwidth]{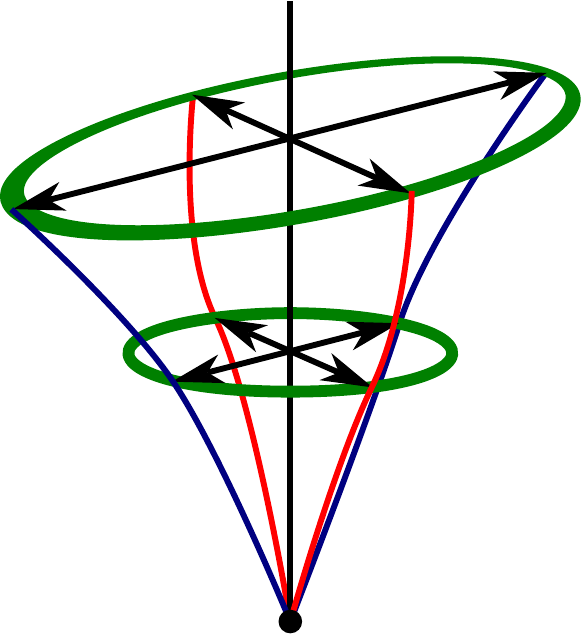}
	\caption[Shear along the polar axis]{Shear along the polar axis. Null geodesics (in red and blue) of constant $ u $ and $ \theta $ emanate from $ r=0 $ (black dot at the bottom) and form an infinitesimal bundle around the axis (vertical black line). Points of constant $r$ (in green) form a set that approximates a circle (below) near the origin. If there is a nonvanishing shear along the axis, this circle is deformed into an ellipse (above). In this case, the vector field \(\partial_{(\theta)}{}^a\) along the axis (black arrows) changes from an initially symmetrical distribution to another one where its `length' has a dependence on $ \phi $.}
	\label{shear}
\end{figure}

The case $\zeta=0$, in which $\gamma|_{\mathcal{Q}_0}=\delta|_{\mathcal{Q}_0}=0$, is the \emph{trivial axial condition}, compatible with the conditions imposed in~\cite{Bondi1962,Isaacson1983} under axial symmetry. It cannot hold in general. If $\zeta$ vanished, then $\partial_r\gamma=\partial_r\delta=0$ along the axis, and the shear scalar \eqref{shear2} would vanish there. This is a coordinate-independent restriction, which cannot be imposed on every spacetime. In general, a bundle of generators around the axis that crosses a region of nonvanishing shear is deformed from a circle into an ellipse \cite[p. 34]{Poisson2004}, so that $g_{\theta\theta}|_{\mathcal{Q}_0}$, $\gamma|_{\mathcal{Q}_0}$, and $\delta|_{\mathcal{Q}_0}$ depend on $\phi$ (figure~\ref{shear}). This ellipse can be described by mapping the coordinate circle $x^2+y^2=\varepsilon^2$, with $\varepsilon\ll1$, from $S^2$ to $\mathcal{S}(u,r)$. Written in terms of proper lengths along the eigenvectors of the matrix of $h_{xx}$, $h_{xy}$, and $h_{yy}$, its image becomes, to leading order in $\varepsilon$, the equation of an ellipse with eccentricity $\sqrt{1-e^{-4|\zeta|}}$ and one principal axis in the direction $\phi_0$.

Following~\cite{Bondi1962}, regularity on the axis can also be expressed by requiring that the ratio between the circumference and the radius of small circles around the axis tend to $2\pi$ as the radius tends to zero. In Bondi--Sachs coordinates, this ratio is $\int_0^{2\pi}e^{-2\gamma}\sech(2\delta)|_{\mathcal{Q}_0}\dd{\phi}+\mathcal{O}(\alpha)$, with $\alpha$ the radius~\cite{Barbosa2026}. By \eqref{ax}, this integral equals $2\pi$ for any $\zeta$ and $\phi_0$. Expanding it in $r$ with the central conditions \eqref{ccg} then gives the \emph{integral axial condition}
\begin{equation}\label{gac}
\int_0^{2 \pi} \left. \gamma^{(2)} \right|_{\mathcal{Q}_0} \dd{\phi} = 0.
\end{equation}

\section{Einstein field equations}\label{EFEs}

\subsection{Hierarchical structure of the solution}

The field equations take a simple form in the coordinates $z^\mu$, which allows them to be solved by a hierarchical integration scheme. Let
\[ H_{ab} \coloneqq G_{ab} + \Lambda g_{ab} - 8\pi T_{ab}, \]
where $G_{ab}$ is the Einstein tensor, $\Lambda$ the cosmological constant, and $T_{ab}$ the energy-momentum tensor. The Einstein field equations, $H_{ab}=0$, imply that
\begin{equation}\label{he}
H_{r\mu} =0,
\end{equation}
\begin{equation}\label{gee}
H_{AA}- \frac{1}{2}g_{AA}g^{BC}H_{BC}=0,
\end{equation}
\begin{equation}\label{mee}
\nabla_\nu H^\nu{}_\mu = -8\pi \nabla_\nu T^\nu{}_\mu =0.
\end{equation}
Equations \eqref{he}, which contain no $\partial_u$, will be called hypersurface equations (HE$\mu$), equations \eqref{gee} gravitational evolution equations (GEE$\theta\theta$ and GEE$\phi\phi$), and equations \eqref{mee} matter evolution equations (MEE$\mu$). Writing $ f(\Psi,\dots) = f(\gamma,\delta,\beta,U,W,\Phi,\dots) $ for functions that may also depend on first- and second-order derivatives of their arguments with respect to non-timelike coordinates, the field equations can be rearranged as
\begin{equation}\label{her}
\qq*{(HE$r$)}  \partial _r \beta = f_\beta(\gamma,\delta,T_{rr}),
\end{equation}
\begin{multline}\label{hetheta}
\qq*{(HE$\theta$)} \partial _r \left(e^{-2 \beta } r^4 \left(e^{2 \gamma } \cosh (2 \delta ) \partial _r U +  \sinh (2 \delta )\sin (\theta )\partial _r W \right)\right) \\
= f_{UW1}(\gamma,\delta,\beta,T_{r\theta}),
\end{multline}
\begin{multline}\label{hephi}
\qq*{(HE$\phi$)} \partial _r \left(e^{-2 \beta } r^4 \left(e^{-2 \gamma }\cosh (2 \delta )  \sin (\theta ) \partial _r W +  \sinh (2 \delta ) \partial _r U \right)\right) \\
= f_{UW2}(\gamma,\delta,\beta,T_{r\phi}),
\end{multline}
\begin{equation}\label{heu}
\qq*{(HE$u$)} \partial _r e^{2 \Phi } + e^{2 \Phi } \frac{ \left(1-4 \pi  r^2 {T}_{{rr}}\right)}{r} = f_\Phi(\gamma,\delta,\beta,U,W,T_{ru},\Lambda),
\end{equation}
\begin{multline}\label{geethetatheta}
\qq*{(GEE$\theta\theta$)} \frac{1}{2} \partial_r \left( r \partial _{u} \gamma \right)  -r \sinh ^2(2 \delta ) \left(\partial _r \gamma \right) \partial _u \gamma   +\\
\tanh (2 \delta ) \left(r \left(\partial _r \delta \right) \partial _u \gamma +\frac{1}{2} \left(\left( 1 +   2 r \partial _r \gamma   \right) \partial _u \delta +r \partial _{{ur}} \delta \right)\right)\\
= f_{\gamma\delta 1}(\Psi,T_{AB}),
\end{multline}
\begin{multline}\label{geephiphi}
\qq*{(GEE$\phi\phi$)} -\frac{1}{2} \partial_r \left( r \partial _{u} \gamma \right) -r \sinh ^2(2 \delta ) \left(\partial _r \gamma \right) \partial _u \gamma    +\\
\tanh (2 \delta ) \left(- r \left(\partial _r \delta \right) \partial _u \gamma + \frac{1}{2} \left( \left(1-2 r \partial _r \gamma \right) \partial _u \delta +r \partial _{{ur}} \delta \right) \right)\\
= f_{\gamma\delta 2}(\Psi,T_{AB}).
\end{multline}
Then, \eqref{geethetatheta} $-$ \eqref{geephiphi} gives equation \eqref{gammaur}, and \eqref{geethetatheta} $+$ \eqref{geephiphi}, divided by $\tanh(2\delta)$, gives equation \eqref{deltaur}:
\begin{equation}\label{gammaur}
\partial_r \left( r \partial _{{u}} \gamma  \right) + 2 \tanh (2 \delta ) \left(\left(\partial _r \delta \right) \left(r \partial _u \gamma  \right)+\left(\partial _r \gamma \right) (r \partial _u \delta )\right) = f^-(\Psi,T_{\theta\theta},T_{\phi\phi}),
\end{equation}
\begin{equation}\label{deltaur}
\partial_r \left( r \partial _{{u}} \delta  \right) - \sinh (4 \delta ) \left(\partial _r \gamma \right) \left( r\partial _u \gamma  \right) = f^+(\Psi,T_{AB}).
\end{equation}
Since the division by $\tanh(2\delta)$ fails where $\delta = 0$, and there the sum of \eqref{geethetatheta} and \eqref{geephiphi} is satisfied identically, equation \eqref{deltaur} needs a separate justification. It is equivalent to $H_{\theta\phi}-\tfrac{1}{2}g_{\theta\phi}g^{BC}H_{BC}=0$, the off-diagonal counterpart of equations \eqref{gee}: the difference between the two sides of \eqref{deltaur} equals $-(e^{2\beta}/2r)\sech(2\delta)\csc(\theta)$ times the left-hand side of this equation. It therefore follows from the Einstein field equations for every $\delta$, including $\delta = 0$.

Moreover, from the matter evolution equations one obtains
\begin{equation}\label{meer}
\qq*{(MEE$r$)} \partial _u T_{rr} = f_{T,r}(\Psi,T_{r\mu},T_{AB}),
\end{equation}
\begin{equation}\label{meetheta}
\qq*{(MEE$\theta$)} \partial _u T_{r\theta} = f_{T,\theta}(\Psi,T_{r\mu},T_{AB},T_{u\theta}),
\end{equation}
\begin{equation}\label{meephi}
\qq*{(MEE$\phi$)} \partial _u T_{r\phi} = f_{T,\phi}(\Psi,T_{r\mu},T_{AB},T_{u\phi}),
\end{equation}
\begin{equation}\label{meeu}
\qq*{(MEE$u$)} \partial _u T_{ru} = -\frac{1}{2} {T}_{{rr}} \partial _u e^{2 \Phi }  + f_{T,u}(\Psi,T_{\mu\nu},\partial_u \gamma,\partial_u \delta,\partial_u \beta,\partial_u U,\partial_u W).
\end{equation}
To solve the last equation, we also take the $u$-derivative of equation \eqref{heu} and substitute $\partial _u T_{ru}$ by the right-hand side of equation \eqref{meeu}, which gives
\begin{equation}\label{Phiur}
\partial _r \left(r \partial _u e^{2 \Phi } \right) = f_{\Phi,u}(\Psi,T_{\mu\nu},\Lambda,\partial_u \gamma,\partial_u \delta,\partial_u \beta,\partial_u U,\partial_u W,\partial_u T_{rr}).
\end{equation}
The explicit forms of the functions $f$ are given in Appendix~\ref{appfs}.

Conversely, a solution of the HEs, equations \eqref{gammaur} and \eqref{deltaur}, and the MEEs satisfies all the Einstein field equations, so that ten equations suffice. With $S_{AB} \coloneqq H_{AB} - \frac{1}{2} g_{AB}\, g^{CD}H_{CD}$, equation \eqref{deltaur} is equivalent to $S_{\theta\phi} = 0$, as shown above, and the difference between the two sides of \eqref{gammaur} is $(e^{2(\beta-\gamma)}/4r)\sech(2\delta)\,(e^{4\gamma}\csc^2(\theta)S_{\phi\phi} - S_{\theta\theta})$. Since $g^{AB}S_{AB} = 0$ identically, \eqref{gammaur} and \eqref{deltaur} together are equivalent to $S_{AB} = 0$, for every $\delta$. These are the gravitational evolution equations of~\cite{Isaacson1983}, and the proof given there applies. In that proof, the components $H_{uA}$ and $H_{uu}$ are found to be proportional to $r^{-2}$, and their vanishing is attributed to smoothness at the origin. A weaker condition can be used: if the metric is $C^2$ and $T_{ab}$ is continuous near $\mathcal{P}$ in the observer's proper coordinates, the components $H_{\bar{\mu}\bar{\nu}}$ are bounded there and, by equations \eqref{x0} and \eqref{xi}, $\partial_u x^{\bar{\mu}} = \mathcal{O}(1)$ and $\partial_A x^{\bar{\mu}} = \mathcal{O}(r)$, so that $H_{uA} = \mathcal{O}(r)$ and $H_{uu} = \mathcal{O}(1)$. Since they are proportional to $r^{-2}$, with coefficients independent of $r$, these bounds can hold only if the coefficients vanish, and so $H_{uA} = H_{uu} = 0$.

A hierarchical integration scheme follows if $\Lambda$ and the 12 functions in $\{ \gamma, \delta, T_{\mu\nu} \}$, independent apart from the constraints listed in Appendix~\ref{firstterms}, are given on an initial light cone $u = u_0$, and if a suitable prescription for the 6 functions in $\{ T_{AB}, T_{uu}, T_{u\theta}, T_{u\phi} \}$ is given for $u > u_0$, for example through equations of motion or equations of state. The scheme proceeds as follows:
\begin{enumerate}
	\item Given the initial data $\left\{ \gamma, \delta, T_{\mu\nu}, \Lambda \right\}$ on $ \mathcal{C}(\tau=u_0) $, integrate equation \eqref{her} for $\beta$ and then the linear system \eqref{hetheta}, \eqref{hephi} for $\partial_r U$ and $\partial_r W$.
	
	\item Integrate the first-order linear differential equation \eqref{heu}, whose solution can be written as integrals of its coefficients. At this point, $\beta$, $U$, $W$, and $\Phi$ are known on the initial light cone.
	
	\item Solve the system of first-order linear differential equations \eqref{gammaur} and \eqref{deltaur} for $r \partial_u  \gamma$ and $r \partial_u  \delta$.
	
	\item Obtain $\partial_u T_{rr}$, $\partial_u T_{r\theta}$, and $\partial_u T_{r\phi}$ directly from equations \eqref{meer}--\eqref{meephi}.
	
	\item Take the $u$-derivative of equations \eqref{her}--\eqref{hephi} and integrate in $r$ as before to obtain $\partial_u \beta$, $\partial_u U$, and $\partial_u W$.
	
	\item Take the $u$-derivative of equation \eqref{heu}, substitute $\partial_u T_{ru}$ by the right-hand side of equation \eqref{meeu} to obtain equation \eqref{Phiur}, and integrate it to find $\partial_u \Phi$.
	
	\item Return to equation \eqref{meeu} to calculate $\partial_u T_{ru}$.
	
	\item With $\partial_u \gamma$, $\partial_u \delta$, $\partial_u T_{r\mu}$, and the suitable prescription for $T_{AB}$, $T_{uu}$, $T_{u\theta}$, and $T_{u\phi}$, either compute numerically the data on the light cone $u = u_0 + \Delta u$ or keep taking $u$-derivatives of the previous equations. Repeating these steps, one either integrates the spacetime numerically for $u>u_0$ or expands all functions to the desired order in $u$.
	
	\item Fix the constants of integration that appear in the process with the central conditions \eqref{ccg} along $\mathcal{P}$.
\end{enumerate}

By a suitable prescription for $T_{AB}$, $T_{uu}$, $T_{u\theta}$, and $T_{u\phi}$ we mean one that fits the scheme: it determines the stresses on each new light cone, while the matter evolution equations evolve the four-momentum $T_{r\mu}$, so that every step above can be carried out. Since each step is either a direct integration in $r$ or the solution of a system of first-order linear differential equations, the data on the initial light cone determine, order by order in $\Delta u = u - u_0$, unique coefficients of the series of all functions, each depending continuously on the data. A solution that is real analytic in $u$ is thus determined, by iteration, within the radius of convergence of this series. This amounts to the `no-interference' condition of~\cite{Ellis1985}: the evolution of the system can be extrapolated from the data on the initial light cone, which excludes, for instance, a lightlike shock wave arriving unannounced along a later light cone.

Whether, for suitable prescriptions, the whole problem, formed by the Einstein field equations and the evolution of matter, is well posed is left to future investigations. A similar question was addressed in~\cite{ChoquetBruhat2011} for data on a single future light cone. There, with the energy-momentum tensor prescribed a priori inside the cone, the Einstein equations in wave-map gauge admit, under certain conditions, a unique solution near the vertex.

The first terms of the expansion in $r$ are given in Appendix~\ref{firstterms}, where one sees that the initial-value constraints of the Cauchy formulation are replaced by much simpler conditions on the coefficients.

The solution has an interesting universal property. The initial data, $\partial_u a^{\bar{i}}$, $\partial_u \Omega^{\bar{i}}$, and the central conditions determine the first $u$-derivatives of $T_{r\mu}$ and of all metric functions without using the prescription for $T_{AB}$, $T_{uu}$, $T_{u\theta}$, and $T_{u\phi}$. Thus, an observer who knows the data on a light cone and how their acceleration and angular velocity will change can compute, to first order in $\Delta u$, the full metric and the four-momentum crossing the next light cone, whatever equations govern the behaviour of matter: its different interactions enter only at second order.

\subsection{Constants of integration}\label{coi}

The constants of integration play a central role in the original Bondi--Sachs formalism \cite{Bondi1962}, where they are related to the mass, the dipole moment, and the news function. Here, they are fixed by the central conditions. Integrating HE$r$ gives
\begin{equation}\label{intbeta}
\beta = \int_{0}^{r} f_{\beta} \dd{r'} + C_{\beta},
\end{equation}
and the central conditions \eqref{ccg} and \eqref{expTBondi} imply that $f_\beta = \mathcal{O}(r)$ and $C_\beta = 0$. Integrating HE$\theta$ and HE$\phi$ once gives
\begin{gather*}
e^{-2 \beta } r^4 (e^{2 \gamma } \cosh (2 \delta ) \partial _r U +  \sinh (2 \delta )\sin (\theta )\partial _r W ) = \int_{0}^{r} f_{UW1} \dd{r'} + C_{UW1},\\
e^{-2 \beta } r^4 (e^{-2 \gamma }\cosh (2 \delta )  \sin (\theta ) \partial _r W +  \sinh (2 \delta ) \partial _r U ) = \int_{0}^{r} f_{UW2} \dd{r'} + C_{UW2}.
\end{gather*}
By the central conditions, the left-hand sides are $\mathcal{O}(r^4)$ and the integrands on the right are $\mathcal{O}(r^3)$, so $C_{UW1} = C_{UW2} = 0$, and a second integration gives
\begin{equation}\label{intU}
U= \int_{0}^{r} \frac{e^{2\beta}}{{r'}^4} \left[ e^{-2\gamma} \cosh(2\delta)\int_{0}^{{r'}} f_{UW1} \dd{r''} - \sinh(2\delta) \int_{0}^{{r'}} f_{UW2} \dd{r''}  \right] \dd{r'} + C_U
\end{equation}
and
\begin{equation}\label{intW}
W= \int_{0}^{r} \frac{e^{2\beta}\csc(\theta)}{{r'}^4} \left[ e^{2\gamma} \cosh(2\delta)\int_{0}^{{r'}} f_{UW2} \dd{r''} - \sinh(2\delta) \int_{0}^{{r'}} f_{UW1} \dd{r''}  \right] \dd{r'} + C_W,
\end{equation}
where the integrands of the outer integrations are $ \mathcal{O} (1)$. Therefore, $ C_U = U^{(0)} $ and $ C_W = W^{(0)} $.
For HE$u$, the integrating factor
\begin{equation}\label{FPhi}
F_\Phi (u,r,\theta,\phi) = \exp(\int_{r^*}^{r} \frac{1}{r'} - 4\pi r' T_{rr} \dd{r'} ), 
\end{equation}
with an arbitrary $r^*>0$, gives
\begin{equation}\label{intPhi}
F_\Phi e^{2\Phi} = \int_{0}^{r} F_\Phi f_\Phi \dd{r'} + C_\Phi, 
\end{equation}
and, since $F_\Phi = \mathcal{O}(r)$ and $F_\Phi f_\Phi = \mathcal{O}(1)$ by the central conditions, both sides are $\mathcal{O}(r)$ only if $C_\Phi = 0$. Finally, the integrating factor of equation \eqref{gammaur} is $\cosh(2\delta)$. With it and the definitions $G\coloneqq r \cosh(2\delta)\partial_u \gamma$ and $D \coloneqq r\partial_u \delta$, equations \eqref{gammaur} and \eqref{deltaur} become the linear system
\begin{equation}\label{GD}
\partial_r G  = -2\sinh(2\delta)\, D\, \partial_r \gamma + \cosh(2\delta)f^- , \qquad \partial_r D  = 2\sinh(2\delta)\, G\, \partial_r \gamma + f^+ .
\end{equation}
Its homogeneous solution is a rotation of $(C_G, C_D)$ by the angle $\omega = \int_0^r 2\sinh(2\delta)\,\partial_{r'}\gamma\, \dd{r'}$, so that, with $H$ the rotation matrix by the angle $\omega$,
\begin{equation}\label{solGD}
\begin{matrix}\mqty(G\\D) = H \mqty(C_G \\ C_D) \end{matrix} + H \int_{0}^{r} H^{-1} \begin{matrix}\mqty(\cosh(2\delta) f^- \\ f^+)\end{matrix} \dd{r'} .
\end{equation}
Since $f^\pm = \mathcal{O}(r)$ and $\omega = \mathcal{O}(r^4)$, $H(r=0)$ is the identity and the integral vanishes as $r\to 0$, so that $C_G = G(r=0) = 0$ and $C_D = D(r=0) = 0$.

Therefore, the only nonvanishing constants of integration are $ C_U = U^{(0)} (u,\theta,\phi) $ and $ C_W = W^{(0)} (u,\theta,\phi) $, in contrast to $ \partial_u c(u,\theta)$, $M(u,\theta)$, and $N(u,\theta)$ in the original formulation \cite{Bondi1962}. They carry the gauge freedom in the choice of the worldline and of its tetrad, through the six functions $a^{\bar{i}}(u)$ and $\Omega^{\bar{i}}(u)$, while $M$ and $N$ are primarily related to the mass and the dipole moment of an isolated system. In particular, the news function $\partial_u c$ has no counterpart: on a light cone, the constants of integration follow from the regularity of the metric at the vertex and, for suitable prescriptions and solutions that are real analytic in $u$, no information enters the domain of integration from outside. Such a counterpart could appear only if the regularity at the vertex were lost, as for a lightlike shock wave propagating along a light cone, across which the transverse extrinsic curvature jumps \cite{Barrabes1991}.

\subsection{Spherically symmetric spacetimes}\label{spherical}

As a simple example of the scheme, we consider spherically symmetric spacetimes, which include, outside a matter distribution, the Schwarzschild--de Sitter spacetime.
Isotropic observations from a centre of symmetry are incompatible with the deformations induced by shear (figure~\ref{shear}), so spherically symmetric spacetimes require vanishing shear, which, by equation \eqref{shear2}, implies that $ \partial_r \gamma=\partial_r \delta=0 $ \cite{Ellis1985}. Together with the central conditions \eqref{ccg}, this implies $ \gamma= \delta=0 $. Then, in analogy with the definition of a static spacetime \cite[p. 119]{Wald1984}, we shall say that the metric \eqref{g} is \textit{nonrotating} if $ \partial_{(u)}{}^a \partial_{(A)}{}_a = 0 $. Combined with the condition on $ \gamma $ and $ \delta $, this leads to $ U=W=0 $, and it follows from the central conditions that $ a^{\bar{i}}=\Omega^{\bar{i}}=0 $.

In addition, there can be no net momentum in angular directions, meaning that $ T_{rA}=g_{ru}g_{AA}T^{uA}=0 $. Shear in the matter content would also imply preferred angular directions and, consequently, we must have $ T_{uA}=g_{uu}g_{AA}T^{uA}+g_{ur}g_{AA}T^{rA}=0 $ and $ T_{\theta\phi}=g_{\theta\theta}g_{\phi\phi}T^{\theta\phi}=0 $. Given that the norm of $ \partial_{(\phi)}{}^a $ depends on $ \theta $, $ T_{\phi\phi} $ may depend on $ \theta $, but not on $ \phi $. The remaining components, which encode the energy density, the radial velocity, and the pressures, must be independent of $ \theta $ and $ \phi $:
\[  T_{uu}=T_{uu}(u,r), \quad T_{ur}=T_{ur}(u,r), \quad T_{rr}=T_{rr}(u,r), \qand  T_{\theta\theta}=T_{\theta\theta}(u,r).\]

Substituting these initial data and $ \Lambda $ into the hypersurface equations \eqref{beta}, \eqref{UW1}, \eqref{UW2}, and \eqref{Phi}, we recover $ U=W=0 $ and find that
\begin{equation}\label{betasphe}
\beta(u,r)= \int_{0}^{r} 2\pi r' T_{rr}(u,r') \dd{r'}  
\end{equation}
and
\begin{equation}\label{phisphe}
\partial_r e^{2\Phi} + \frac{1-4\pi r^2 T_{rr}}{r}e^{2\Phi} = \frac{1-r^2 \Lambda}{r}e^{2\beta}+ 8\pi r T_{ur}.
\end{equation}
Multiplying the last equation by the integrating factor \eqref{FPhi} and integrating yields
\[ e^{2\Phi}= F_\Phi^{-1}\int_{0}^{r} F_\Phi \left(\frac{1-{r'}^2 \Lambda}{r'}e^{2\beta}+ 8\pi r' T_{ur}\right) \dd{r'}.\]
From \eqref{betasphe}, one finds that
\[ F_\Phi(u,r)= \frac{r}{r^*}e^{2\beta(u,r^*)}e^{-2\beta(u,r)} \]
and
\begin{equation}\label{e2P}
e^{2\Phi} = e^{2\beta}\left(  1-\frac{\Lambda}{3}r^2+\frac{1}{r}   \int_{0}^{r} 8\pi {r'}^2 T_{ur} e^{-2\beta} \dd{r'} \right).
\end{equation}

Then, the metric \eqref{g} becomes
\begin{equation}\label{gsphe}
\dd{s}^2=  -e^{2(\Phi+\beta)} \dd{u}^2 +2 e^{2\beta}\dd{u}\dd{r}+r^2\dd{\theta}^2+ r^2\sin[2](\theta)\dd{\phi}^2. 
\end{equation}
The nontrivial initial-value constraints in Appendix~\ref{firstterms} that result from the restrictions on $ T_{\mu\nu}$, $\gamma$, and $ \delta $ are
\begin{equation}\label{ivc}
T_{\theta\theta}{}^{(2)}=\frac{T_{\phi\phi}{}^{(2)}}{\sin^2(\theta)}=2\left(T_{rr}{}^{(0)}+T_{ur}{}^{(0)}\right), \qquad T_{uu}{}^{(0)}=-T_{ur}{}^{(0)}.
\end{equation}

If the matter is confined within a radius $ R $, it is useful to define
\begin{equation}\label{M}
M(u,R)\coloneqq -\int_{0}^{R} 4\pi r^2 T_{ur}(u,r)e^{-2\beta(u,r)} \dd{r},
\end{equation}
a quasi-local, past-light-cone version of the Bondi mass in equation (5.22) of~\cite{Siebel2002}, and
\begin{equation}\label{betaR}
\beta_R(u)\coloneqq\beta(u,R)= \int_{0}^{R} 2\pi r T_{rr}(u,r) \dd{r}  .
\end{equation}
Then, for $ r\geq R $,
\begin{equation}\label{e2phi}
e^{2\Phi} = e^{2\beta_R}\left(  1-\frac{\Lambda}{3}r^2-\frac{2M}{r}   \right)
\end{equation}
and
\[ \Phi= \beta_R + \frac{1}{2}\ln\left(  1-\frac{\Lambda}{3}r^2-\frac{2M}{r}  \right). \]

Expanding the logarithm to first order, one recovers the Newtonian gravitational potential,
\[ \Phi\approx-\frac{1}{r}\left(M +\frac{4\pi r^3}{3}\frac{\Lambda}{8\pi }\right)+ \text{constant}, \]
with an extra mass term due to the cosmological constant. Moreover, the dominant energy condition \cite{Poisson2004} implies that $ -T^a{}_b \partial_{(u)}{}^b $ is a future-directed timelike or null vector field (as long as the coordinate system remains well defined and $ \partial_{(u)}{}^a $ remains timelike). Since $ \partial_{(r)}{}^a $ is past-directed and null, it follows that $ -T_{ab} \partial_{(u)}{}^a \partial_{(r)}{}^b = -T_{ur}\geq 0 $, and, from definition \eqref{M}, that
\begin{equation}\label{Mpositive}
M\geq0,
\end{equation}
with equality holding only for $ T_{ab}=0 $ or $ T^a{}_b \partial_{(u)}{}^b \propto \partial_{(r)}{}^a$. For constant $ \beta_R $ and $ M $, the geodesic equation with zero initial velocity gives
\begin{equation}\label{force}
\dv[2]{r}{u} =-\frac{\partial_{r}e^{4\Phi}}{4}=  -e^{4\Phi}\partial_{r}\Phi.
\end{equation}
In a weak-field approximation with $ \Phi \ll 1 $, it follows that
\[\dv[2]{r}{u}  \approx-\partial_{r}\Phi. \]
The mass $ M $ may therefore be regarded as a generalisation of the Newtonian active gravitational mass of the matter content, and $ \Phi $ as a generalisation of the Newtonian gravitational potential.

When $ \beta_R $ and $ M $ do not depend on $ u $ and $ \Lambda > 0 $, this is the Schwarzschild--de Sitter spacetime.

The formalism can also serve as a guide when $T_{\mu\nu}$ is not prescribed beforehand. In Appendix \ref{dustapp}, we consider spherically symmetric spacetimes filled with dust, whose stresses follow from its equation of state, and adapt the order of the scheme accordingly. Both examples are used in the analysis of the quasi-local emission of Hawking radiation in~\cite{Barbosa2026}.

\section{Analogue of the integral mass formula}\label{imf}

We now identify the mass content seen by an observer immersed in a general matter distribution and relate it to the integral mass formula of black hole mechanics \cite{Bardeen1973}.

\subsection{Local mean density}

In the Newtonian limit, $g_{uu} \simeq -1 - 2\Phi_N$, where $\Phi_N$ is the gravitational potential \cite[p. 415]{Misner1973}. We define $\Phi_N \coloneqq (-1-g_{uu})/2$ in general and expand it in $r$:
\[ \Phi_N = \Phi ^{{(1)}} r + \frac{1}{2} \left( \Phi ^{(2)} + 2 ( \Phi ^{(1)} )^2+\beta ^{(2)} - (U^{{(0)}} )^2-\sin ^2(\theta ) ( W^{{(0)}} )^2\right) r^2 +\mathcal{O}\left(r^3\right). \]
The linear term represents inertial forces \cite[p. 295]{Misner1973}, and the terms $\mathcal{O}(r^3)$ describe how the mass is distributed around the observer, much as multipole moments do. For a Newtonian density $\rho = \rho_0 + \mathcal{O}(r)$, Poisson's equation gives $\Phi_N = (2\pi/3)\rho_0 r^2 + \mathcal{O}(r^3)$ \cite[p. 28]{PoissonWill2014}, so we define the mass density at the observer as
\[ \rho \coloneqq  \frac{3\Phi_N^{(2)}}{4\pi}. \]	
The first terms of the solution (Appendix~\ref{firstterms}) give $\rho$ in terms of $T_{ur}{}^{(0)}$, $T_{rr}{}^{(0)}$, $T_{r\theta}{}^{(1)}$, $T_{r\phi}{}^{(1)}$, $\gamma^{(2)}$, $\delta^{(2)}$, $\Lambda$, $U^{(0)}$, and $W^{(0)}$, and the central conditions \eqref{expTBondi} then give
\begin{equation}\label{rhobar}
\begin{aligned}
\rho = & \; {T}^{\bar{0}\bar{0}{(0)}}+\frac{1}{4}  \left(5+3 \cos (2 \theta )-6 \sin ^2(\theta ) \cos (2 \phi )\right){T}^{\bar{1}\bar{1}{(0)}} +\\
&\; \frac{1}{4}  \left(5+3 \cos (2 \theta ) + 6 \sin ^2(\theta ) \cos (2 \phi )\right){T}^{\bar{2}\bar{2}{(0)}}+\frac{1}{2} (1-3 \cos (2 \theta )) {T}^{\bar{3}\bar{3}{(0)}} 	-\\
&\; 3 \left(\sin ^2(\theta ) \sin (2 \phi ){T}^{\bar{1}\bar{2}{(0)}} +\sin (2 \theta ) \left( \cos (\phi ) {T}^{\bar{1}\bar{3}{(0)}} + \sin (\phi ){T}^{\bar{2}\bar{3}{(0)}}\right)\right) +\\
&\; \frac{3}{16 \pi } \left(2 \gamma ^{{(2)}} + \csc ^2(\theta ) \partial _{\phi \phi } \gamma ^{{(2)}} -3 \cot (\theta ) \partial _{\theta } \gamma ^{{(2)}} -\partial _{\theta \theta } \gamma ^{{(2)}}-2 \csc (\theta ) \partial _{\theta \phi } \delta ^{{(2)}} - \right .\\
&\; \left .2 \cot (\theta ) \csc (\theta ) \partial _{\phi } \delta ^{{(2)}} \right)-\frac{\Lambda }{4 \pi }  -\frac{3}{4 \pi } \left(( U^{{(0)}} )^2+\sin ^2(\theta ) ( W^{{(0)}} )^2\right) .
\end{aligned}
\end{equation}

To remove the angular dependence and obtain the monopole term of the source of the gravitational potential, we take a mean over the light cone. An integration over the angular coordinates alone would suffice, as in the original formulation \cite{Bondi1962}, but a three-dimensional mean is better suited to the notion of a thermodynamic system contained in a light cone, in the spirit of the `photographic' thermodynamics of~\cite{Dunkel2009}, where the system is what the observer sees on their past light cone. However, a null hypersurface has no canonical scalar surface element. In Bondi--Sachs coordinates, the vector surface element compatible with Stokes' theorem \cite{Poisson2004,Wald1984} is $\dd{\Sigma_\mu} = \partial_{(r)}{}_{\mu}\, r^2 \sin(\theta) \dd{r} \dd{\theta} \dd{\phi}$ on a light cone, and a transversal vector must be chosen to project it, a role played by the unit normal on non-null hypersurfaces. We choose $\partial_{(u)}{}^a$, which gives
\begin{equation}\label{dSigma}
\dd{\Sigma} \coloneqq  \partial_{(u)}{}^{\mu} \dd{\Sigma_\mu} = e^{2\beta}r^2 \sin(\theta) \dd{r} \dd{\theta} \dd{\phi}.
\end{equation}
This choice mimics Newtonian physics in two ways. First, the four-dimensional volume element splits as $\dd{v} = \dd{u}\dd{\Sigma}$, so that $u$ carries a unit integration measure, like the absolute time of Newtonian mechanics. Second, by equation \eqref{delrlambda}, $\dd{\Sigma} = ( \sqrt{\det(g_{A'B'})}\dd{\theta'} \dd{\phi'} ) \dd{\lambda'}$, so that the affine parameter, whose tangent vector $k^a$ is calibrated by the observer along $\mathcal{P}$ and parallelly transported, also carries a unit measure, as linear coordinates do in Newtonian theory. In this sense, $\dd{\Sigma}$ is the `most Newtonian' scalar surface element for a light cone, in which the differences between Newtonian theory and general relativity are concentrated in $\det(g_{A'B'})$.

With this surface element, the mean mass density inside a ball $ \mathcal{B}_R $ of radius $ r=R $ and volume $ V_R $, $\bar{\rho} (u,R) \coloneqq V_R^{-1} \int_{\mathcal{B}_R} \rho \dd{\Sigma}$, defines the \emph{local mean density}
\[\bar{\rho}^{(0)} \coloneqq  \bar{\rho} (u,0) = \lim_{R\to 0} \frac{3}{4\pi R^3} \int_{\mathcal{B}_R} \rho r^2 \sin(\theta) \dd{r} \dd{\theta} \dd{\phi}  + \mathcal{O}(R^2) = \frac{1}{4\pi} \int_{S^2} \rho \sin(\theta)  \dd{\theta} \dd{\phi}. \]

\subsection{The mass formula}

To make the analogy with the integral mass formula apparent, we use the kinematics of null hypersurfaces \cite{Gourgoulhon2006}. There is a preferred transversal null vector field,
\begin{equation}\label{ldef}
l^a, \qq{such that} l_al^a=0, \quad l_a \partial_{(r)}{}^a=1 \qand l_a \partial_{(A)}{}^a=0,
\end{equation}
with which we define the projector onto the surfaces $ \mathcal{S}(u,r) $, $ q^{a}{}_{b} \coloneqq \delta^{a}{}_{b} - \partial_{(r)}{}^a l_b- l^a \partial_{(r)}{}_b$, and the normal curvature $ \Theta_{ab} $, the expansion scalar $ \Theta $, the shear tensor $ \sigma_{ab} $, the H\'aj\'i\v{c}ek one-form $ \Omega^H{}_a $, the transversal deformation rate $ \Xi_{ab} $, the transversal expansion scalar $ \Xi $, and the inaffinity $\kappa$:
\[ \Theta_{ab} \coloneqq  q^{c}{}_{a}q^{d}{}_{b} \nabla_c \partial_{(r)}{}_d, \quad \Theta \coloneqq g^{ab} \Theta_{ab} = \frac{2}{r}, \quad \sigma_{ab} \coloneqq \Theta_{ab} - \frac{\Theta}{2} q_{ab}, \]
\[ \Omega^H{}_a \coloneqq  q^{c}{}_{a}l^{d} \nabla_c \partial_{(r)}{}_d, \quad \Xi_{ab} \coloneqq q^{c}{}_{a}q^{d}{}_{b} \nabla_c l_d , \quad \Xi \coloneqq g^{ab} \Xi_{ab},\]
\begin{equation}\label{kappa}
\kappa \coloneqq \partial_{(r)}{}^a l^b \nabla_a \partial_{(r)}{}_b .
\end{equation}
We also introduce the \emph{reduced transversal expansion scalar},
\begin{equation}\label{xidef}
\xi \coloneqq 2\Xi - \Theta,
\end{equation}
which removes the trivial divergence of $ \Xi $. Writing $T_{ur}{}^{(0)}$, $T_{rr}{}^{(0)}$, $T_{r\theta}{}^{(1)}$, and $T_{r\phi}{}^{(1)}$ in terms of these quantities (Appendix \ref{kinematics}) and substituting them into the expression for $\rho$ given by the first terms of the solution, before the central conditions \eqref{expTBondi} are imposed, gives
\begin{equation}\label{rhokine}
\begin{aligned}
\frac{\rho}{3} = &\;\frac{(\kappa\Theta)^{(0)} + \xi^{(1)}}{8\pi} + \frac{\partial_\theta\left(\sin(\theta)\,\Omega^H{}_\theta{}^{(2)}\right) + \csc(\theta)\,\partial_\phi\Omega^H{}_\phi{}^{(2)}}{4\pi\sin(\theta)}\\
&-\frac{1}{2}\left[\frac{\partial_\theta\left(\sin(\theta)\,\partial_\theta T_{rr}{}^{(0)}\right)}{\sin(\theta)} + \frac{\partial_{\phi\phi}T_{rr}{}^{(0)}}{\sin^2(\theta)}\right] - \frac{(U^{(0)})^2 + \sin^2(\theta)\,(W^{(0)})^2}{4\pi}.
\end{aligned}
\end{equation}
In the angular mean, the terms with $U^{(0)}$ and $W^{(0)}$ give, by the central conditions \eqref{ccg}, a term depending on $ a^2 = a_{\bar{i}} a^{\bar{i}}$ and $ \Omega^2 = \Omega_{\bar{i}} \Omega^{\bar{i}}$, the terms with $T_{rr}{}^{(0)}$ integrate to zero by conditions \eqref{expTBondi}, and the terms with the H\'aj\'i\v{c}ek one-form integrate to boundary terms:
\begin{equation}\label{lmd}
\begin{aligned}
\frac{4\pi}{3}\bar{\rho}^{(0)} = &\; \int_{S^2} \frac{(\kappa \Theta) ^{{(0)}}+\xi ^{(1)}}{8 \pi } \sin(\theta) \dd{\theta} \dd{\phi} -\frac{4\pi}{3}\frac{a^2 +\Omega^2}{2\pi}+\\
&\; \int_{0}^{2\pi} \left(\left. \frac{\sin(\theta)\Omega ^H{}_{\theta }{}^{{(2)}}}{4\pi}\right|_{\theta=0}^{\theta=\pi}\right) \dd{\phi} + \int_{0}^{\pi} \left(\left. \frac{\Omega ^H{}_{\phi }{}^{{(2)}}}{4\pi\sin(\theta)}\right|_{\phi=0}^{\phi=2\pi}\right) \dd{\theta}.
\end{aligned}
\end{equation}
The integrand of the last integral vanishes identically, and the other boundary term vanishes by equation \eqref{Omegatheta} with the central conditions \eqref{expTBondi}, the finiteness of $ \partial_\theta \gamma^{(2)} $ on the axis and the integral axial condition \eqref{gac}. Hence,
\begin{equation}\label{lmdfinal}
\frac{4\pi}{3}\bar{\rho}^{(0)} = \int_{S^2} \frac{(\kappa \Theta) ^{{(0)}}+\xi ^{(1)}}{8 \pi } \sin(\theta) \dd{\theta} \dd{\phi} +\frac{4\pi}{3}\rho_{\mathrm{fic}}, \qquad \rho_{\mathrm{fic}} \coloneqq - \frac{a^2 +\Omega^2}{2\pi},
\end{equation}
where we call $\rho_{\mathrm{fic}}$ the \emph{fictitious mass density}.

Equation \eqref{lmdfinal} relates the local mean density to the angular integral of two coefficients of the expansion at the vertex, one associated with the generators of the cone, $(\kappa\Theta)^{(0)}$, and one with the transversal null direction, $\xi^{(1)}$, apart from the contribution of the observer's acceleration and rotation. It has a structure similar to that of equation~(8) of Box~17.2 of~\cite{Misner1973}, a hypersurface constraint that equates the sum of the intrinsic and extrinsic curvature scalars of a spacelike hypersurface to $16\pi$ times the local density of mass-energy. In general, the two differ: here the hypersurface is null, and the source is the active mass density, as shown below, since $\rho$ was defined from the potential $\Phi_N$. In the Newtonian limit, however, with negligible stresses and cosmological constant and an inertial observer ($a = \Omega = 0$), both sources reduce to the energy density ${T}^{\bar{0}\bar{0}{(0)}}$. Equation \eqref{lmdfinal} is also the counterpart, at the vertex, of the expression of~\cite{Chrusciel2014} for the Trautman--Bondi mass of a complete future light cone, $m_{\mathrm{TB}} = (16\pi)^{-1}\int_{S^2}(\zeta_2+\tau_2)\dd{\mu_s}$, where $\tau_2$ and $\zeta_2$ are the coefficients of $r^{-2}$ in the expansions, for large $r$, of the expansion of the generators and of the divergence of the transversal null normal. In both cases, a mass is read off from the expansions of the tangent and transversal null directions, near null infinity in one case and at the observer in the other.

Most importantly, equation \eqref{lmdfinal} has the structure of the integral mass formula for a nonrotating black hole \cite{Bardeen1973}, which writes the mass as the sum of an area term and a matter term. The matter term is an integral, over a hypersurface extending outward from the horizon, of $T_{ab}-\frac{1}{2}T g_{ab}$ contracted with the unit normal of this hypersurface and with the stationary Killing field. The first term of equation \eqref{lmdfinal} corresponds to the area term, since the expansion scalar measures the rate of change in area of the surfaces $\mathcal{S}(u,r)$ along the generators. The second, built from the transversal null vector $l^a$, corresponds to the matter term. On the horizon, the Killing field is the null generator and hence also the normal of the horizon, so the contraction with it picks out the components of the energy-momentum tensor transversal to the horizon. Similarly, by equation \eqref{xi1}, $\xi^{(1)}$ contains $T_{ur}{}^{(0)} = T_{ab}\,\partial_{(u)}{}^a\partial_{(r)}{}^b$, in which $\partial_{(r)}{}^a$ is the normal of the light cone, and $\partial_{(u)}{}^a$, which defines the surface element \eqref{dSigma}, plays the role of the unit normal of the hypersurface of integration. To make the correspondence between the area terms precise, let
\[ Q \coloneqq \frac{3}{8\pi} \int_{\mathcal{B}_R} \kappa \Theta \dd{\Sigma}. \]
Using equations \eqref{dSigma}, \eqref{kappadelrbeta}, and \eqref{Theta}, and expanding with the central conditions \eqref{ccg},
\begin{equation*}
Q = \frac{3}{2\pi} \int_{S^2}\int_{0}^{R} (\partial_r \beta) e^{2\beta} r \sin(\theta) \dd{r} \dd{\theta} \dd{\phi} = \frac{R^2}{4\pi} \int_{S^2} 2\beta^{(2)}R\sin(\theta) \dd{\theta} \dd{\phi}  + \mathcal{O}(R^4).
\end{equation*}
With the expansion $\kappa = 2\beta^{(2)} r + \mathcal{O}(r^2)$ and the average per solid angle
\[ \hat{\kappa} (u,r) \coloneqq \frac{1}{4\pi} \int_{S^2} \kappa(u,r,\theta,\phi) \sin(\theta) \dd{\theta} \dd{\phi}  , \]
one finds $ Q = \hat{\kappa}(u,R)A_R/4\pi  + \mathcal{O}(R^4)$, where $A_R = 4\pi R^2$ is the area of $\mathcal{S}(u,R)$. The associated density $ \bar{\rho}_\parallel \coloneqq Q/V_R $ then satisfies
\begin{equation}\label{rhoparallel}
\frac{4\pi}{3}\bar{\rho}_\parallel{}^{(0)} = \int_{S^2} \frac{(\kappa \Theta) ^{{(0)}} }{8 \pi } \sin(\theta)\dd{\theta} \dd{\phi}  = \lim_{R\to 0} \frac{\hat{\kappa}(u,R)A_R}{4\pi}   \frac{1}{R^3}.
\end{equation}
With
\[ \frac{4\pi}{3}\bar{\rho}_\perp{}^{(0)}\coloneqq  \int_{S^2} \frac{\xi ^{(1)}}{8 \pi } \sin(\theta) \dd{\theta} \dd{\phi}, \]
the total mass inside an infinitesimal ball around the observer is
\begin{equation}\label{rhosplit}
\frac{4\pi}{3}\bar{\rho}^{(0)} \dd{r}^3 = \frac{4\pi}{3} \left( \bar{\rho}_\parallel{}^{(0)}  + \bar{\rho}_\perp{}^{(0)} +  \rho_{\mathrm{fic}} \right) \dd{r}^3 .
\end{equation}
Expanding the limit in equation \eqref{rhoparallel} as
\[ \frac{\hat{\kappa}(u,r)A_r}{4\pi r^3} = \lim_{R\to 0} \frac{\hat{\kappa}(u,R)A_R}{4\pi R^3} + \mathcal{O}(r), \]
we define the \emph{infinitesimal gravitational mass} as the lowest-order term of this mass:
\begin{equation}\label{igm}
M_{\dd{r}}(u) \coloneqq \frac{\hat{\kappa}(u,\dd{r})A_{\dd{r}}}{4\pi}  +\frac{4\pi}{3} \left( \bar{\rho}_\perp{}^{(0)} +  \rho_{\mathrm{fic}} \right) \dd{r}^3.
\end{equation}
This is the analogue of the integral mass formula for a nonrotating black hole. Both are related to the leading term in a multipole expansion of a Newtonian-like potential built from the metric, so both can be interpreted as the gravitational mass of the system, and both split into a term arising from a vector field tangent to the causal horizon and a term related to transversal structures. Remarkably, the area term has exactly the same form in both cases, including the numerical factor.

In terms of the energy-momentum tensor, equation \eqref{xi1} shows that the H\'aj\'i\v{c}ek terms in $\bar{\rho}_\perp{}^{(0)}$ are $-1/6$ times the boundary terms of equation \eqref{lmd}, so they vanish by the same arguments. The remaining terms with $\gamma^{(2)}$ and $\delta^{(2)}$ are total derivatives in $\phi$ and $\theta$, which integrate to zero by periodicity and, as in equation \eqref{lmd}, by the finiteness of $ \partial_\theta \gamma^{(2)} $ on the axis and the integral axial condition \eqref{gac}. The central conditions \eqref{expTBondi} then give
\[ \frac{4\pi}{3}\bar{\rho}_\perp{}^{(0)} = -\frac{8\pi}{3}\left( {T}^{\bar{0}\bar{0}{(0)}} + \frac{\Lambda}{8\pi}\right), \]
while the area term, by equation \eqref{kT0} and the central conditions \eqref{expTBondi}, is
\begin{equation}\label{parallel}
\frac{4\pi}{3}\bar{\rho}_\parallel{}^{(0)} = \frac{4\pi}{3} \left( 3{T}^{\bar{0}\bar{0}{(0)}} + {T}^{\bar{1}\bar{1}{(0)}} + {T}^{\bar{2}\bar{2}{(0)}} + {T}^{\bar{3}\bar{3}{(0)}}\right).
\end{equation}
Therefore,
\begin{equation}\label{igmexplicit}
M_{\dd{r}}(u) = \frac{4\pi}{3}\left[ {T}^{\bar{0}\bar{0}{(0)}} + \frac{\Lambda}{8\pi} - \frac{a^2}{2\pi} + \sum_{\bar{i}} \left( {T}^{\bar{i}\bar{i}{(0)}} - \frac{\Lambda}{8\pi} - \frac{\left(\Omega^{\bar{i}}\right)^2}{2\pi} \right) \right] \dd{r}^3 .
\end{equation}
The contribution of $\Lambda$ is that of a fluid with energy density $\Lambda/8\pi$ and pressure $-\Lambda/8\pi$. Curiously, the fictitious density splits in a similar way: the acceleration enters like an energy density, and each component of the angular velocity like the pressure along that direction.

For a perfect fluid at rest relative to the observer, with energy density $ \mu $ and pressure $ p $, the matter term is $ \mu+3p $. It is the source of the Newtonian-like potential built from the metric in the interior Schwarzschild solution \cite[p. 127]{Wald1984}, which is the counterpart of $\Phi_N$, and the integrand of the Komar mass of static spacetimes \cite{Komar1959,Wald1984}, on which the integral mass formula of black holes is based \cite{Bardeen1973}.

\section{Conclusions}\label{conclusions}

We have generalised the Bondi--Sachs formalism to the past light cones of an arbitrary timelike worldline, for an accelerated observer with a rotating tetrad, in the presence of matter and a cosmological constant, and without symmetry assumptions. We illustrated it in spherically symmetric spacetimes, which include the Schwarzschild--de Sitter spacetime. The central conditions, for the metric and for the energy-momentum tensor, fix all constants of integration: the only nonvanishing ones carry the gauge freedom of the worldline and its tetrad, and there is no analogue of the news function. The general axial conditions accommodate shear along the axis, where the trivial condition used under axial symmetry fails. The field equations can then be integrated hierarchically from data on a single light cone, and, to first order in the observer's proper time, the evolution of the metric and of the four-momentum crossing the light cones is independent of the specific behaviour of matter.

As an application, we derived an analogue of the integral mass formula of black hole mechanics for an observer immersed in matter. The mass associated with the monopole of a Newtonian-like potential built from the metric splits into an area term, with the same form and numerical factor as for a nonrotating black hole, a term built from transversal structures, and a fictitious contribution fixed by the observer's acceleration and rotation. Its matter content reduces to $\mu+3p$ for a perfect fluid, as does the density of the Komar mass of static spacetimes.

These results support the proposal of~\cite{Barbosa2026} that past light cones of arbitrary worldlines provide a natural setting for the thermodynamic aspects of general relativity. There, the area term of the mass formula is related to a temperature and an entropy, and the formalism is used to analyse the quasi-local emission of Hawking radiation, an analogue of the second law of thermodynamics, and the internal energy of spacetime. Beyond this programme, the generality of the formalism may be useful for both analytic and numerical calculations in a wide range of applications. A numerical implementation of the hierarchical scheme is a natural next step. However, the question of whether general classes of matter content can be incorporated into this formalism in a way that renders a well-posed problem remains open.

\bmhead{Acknowledgements}

The author would like to thank Professor Daniel A. T. Vanzella for presenting him with a fascinating problem, providing valuable feedback, and being supportive during this project. He also thanks Professor Lucas C. Céleri for helpful references and comments.

\bmhead{Statements and Declarations}

\begin{itemize}
\item \textbf{Data accessibility.} This article has no additional data.
\item \textbf{Declaration of AI use.} Claude Opus 5.5 and Claude Fable 5.2 (Anthropic) were used to condense the author's thesis into this article, draft the abstract, introduction, and conclusions, reformulate the derivation of the axial conditions, edit the language, suggest references, and independently verify the equations with numerical scripts, as described in section~\ref{intro}. The author reviewed the whole text and takes full responsibility for its content.
\item \textbf{Competing interests.} I declare I have no competing interests.
\item \textbf{Funding.} This study was financed in part by the Coordenação de Aperfeiçoamento de Pessoal de Nível Superior -- Brasil (CAPES) -- Finance Code 001.
\end{itemize}

\begin{appendices}
\renewcommand{\theequation}{\thesection\arabic{equation}}

\section{Spherically symmetric dusty spacetimes}\label{dustapp}

In this appendix, we adapt the scheme, using the general results of section~\ref{spherical}, to a spherically symmetric spacetime filled with dust, a pressureless perfect fluid with proper energy density $ \mu(u,r) $. Spherical symmetry requires that its four-velocity, $ v^a $, be orthogonal to $ \partial_{(\theta)}{}^a $ and $ \partial_{(\phi)}{}^a $. Then, $ T_{\theta\theta}=T_{\phi\phi}=0 $ and
\[ T_{ab}= \mu(u,r)v_av_b.\]
Since $ v^av_a=-1 $, contracting the last equality with the inverse metric \eqref{g} leads to
\begin{equation}\label{Tur}
T_{ur} = -\frac{e^{2\beta}}{2}\mu  -\frac{e^{2\Phi}}{2}T_{rr}.
\end{equation}

Following~\cite{Ellis1985}, we write the four-velocity of the fluid in terms of the observed redshift $z$ as $v^{u'}=1+z(u',\lambda')$. The coordinate transformation \eqref{ytoz} then gives
\begin{equation}\label{vu}
v^u=1+z(u,r)>0,
\end{equation}
where the inequality is required for a future-directed, timelike flow of dust. The normalisation of this vector field implies that
\begin{equation}\label{vr}
 v^r=\frac{e^{2\Phi}(1+z)}{2}-\frac{e^{-2\beta}}{2(1+z)},
\end{equation}
and its covariant components are
\[v_u=-\frac{1+e^{2(\Phi+\beta)}(1+z)^2}{2(1+z)} \qand v_r= e^{2\beta}(1+z).\]
Therefore,
\begin{equation}\label{Trr}
T_{rr}=\mu e^{4\beta}(1+z)^2
\end{equation}
and
\begin{equation}\label{Tuu}
T_{uu}= \frac{\mu}{4(1+z)^2} \left[1+e^{2(\Phi+\beta)}(1+z)^2\right]^2 = \frac{e^{-4\beta}}{\mu(1+z)^2}T_{ur}{}^2.
\end{equation}

In this model, $ T_{\mu\nu} $ is not prescribed beforehand, and the order of the hierarchical scheme must be adapted. Equation \eqref{betasphe} becomes
\[ \partial_r \beta= 2\pi r \mu (1+z)^2 e^{4\beta}, \]
and can be solved by defining $ x\coloneqq e^{4\beta} $. This yields a Bernoulli differential equation,
\[ \partial_r x= 8\pi r \mu (1+z)^2 x^2, \]
which is reduced to a linear differential equation by $ y\coloneqq x^{-1} $. Integrating and substituting back gives
\begin{equation}\label{betasol}
\beta=-\frac{1}{4} \ln\left( 1- \int_{0}^{r} 8\pi r' \mu (1+z)^2 \dd{r'}   \right).
\end{equation}
Inserting this into equation \eqref{Trr} gives
\begin{equation}\label{Trrsol}
T_{rr} = \frac{\mu(1+z)^2}{1- \int_{0}^{r} 8\pi r' \mu (1+z)^2 \dd{r'}}. 
\end{equation}

Then, multiplying equation \eqref{phisphe} by $ r $ and using equation \eqref{Tur} leads to
\begin{equation}\label{delrre2Phi}
\partial_r \left(r e^{2\Phi}\right)= e^{2\beta}\left[1-(4\pi \mu + \Lambda)r^2\right],
\end{equation}
which, together with equation \eqref{betasol}, results in
\begin{equation}\label{phisol}
e^{2\Phi} = \frac{1}{r} \int_{0}^{r}\frac{1-(4\pi \mu + \Lambda){r'}^2}{\sqrt{1- \int_{0}^{r'} 8\pi r'' \mu (1+z)^2 \dd{r''}}} \dd{r'}.
\end{equation}
Finally, one can compute $ T_{ur} $ from equations \eqref{Tur}, \eqref{betasol}, \eqref{phisol}, and \eqref{Trrsol}:
\begin{multline}\label{Tursol}
T_{ur}= -\frac{\mu}{2\sqrt{1- \int_{0}^{r} 8\pi r' \mu (1+z)^2 \dd{r'}}} -\\
\frac{\mu(1+z)^2}{2r\left[1- \int_{0}^{r} 8\pi r' \mu (1+z)^2 \dd{r'}\right]} \int_{0}^{r}\frac{1-(4\pi \mu + \Lambda){r'}^2}{\sqrt{1- \int_{0}^{r'} 8\pi r'' \mu (1+z)^2 \dd{r''}}} \dd{r'}.
\end{multline}

The time evolution of this spacetime is given by equation \eqref{Phiur} and by equations \eqref{deluTrr} and \eqref{deluTur}. After inserting the components of the energy-momentum tensor as given above, expressing $ \partial_r \beta $ and $ \partial_r \Phi $ using equations \eqref{betasphe} and \eqref{phisphe} and using equations \eqref{Tur} and \eqref{Tuu}, one obtains
\begin{equation}\label{deluTrrsphe}
\partial_u T_{rr} = \frac{e^{2\beta}}{2r^2}\partial_r(r^2\mu) -\frac{e^{2\Phi}}{2}\partial_r T_{rr} - \frac{e^{2\beta}}{r}T_{rr}\left[1-(4\pi \mu+\Lambda)r^2\right].
\end{equation}
Using this result and equations \eqref{Trr} and \eqref{vr} in equation \eqref{Phiur}, one finds
\begin{equation}\label{delue2Phi}
\partial_u e^{2\Phi} = \frac{1}{r}\int_{0}^{r} \frac{4\pi}{1+z} \left[\frac{{r'}^2\mu\partial_{r'} z}{(1+z)^2}+e^{2\beta}v^r\partial_{r'}({r'}^2\mu)\right]+ \left[1-(4\pi\mu+\Lambda){r'}^2\right]\partial_u e^{2\beta} \dd{r'},
\end{equation}
where $ \partial_u \beta $ can be computed by taking the $ u $-derivative of equation \eqref{betasphe} and using \eqref{deluTrrsphe}.

The last equation then gives $ \partial_u T_{ur} $, from which equation \eqref{Tur} yields  
\begin{equation}\label{delumu}
\partial_u \mu=-\frac{1}{1+z}\left[r^{-2}v^r\partial_r\left(r^2\mu\right)+\frac{\mu e^{-2\beta}}{(1+z)^2}\partial_r z\right] .
\end{equation}
Finally, $ \partial_u z $ can be calculated using equation \eqref{Trr}.

The initial-value constraints can be evaluated by expanding $ \mu $ and $ z $ as in equation \eqref{f}. Using the central conditions \eqref{ccg} and equations \eqref{Trr}, \eqref{Tuu}, and \eqref{Tur}, all three constraints \eqref{ivc} become equivalent to
\begin{equation}\label{ivcTrrTur}
T_{rr}{}^{(0)}=-T_{ur}{}^{(0)},
\end{equation}
which is satisfied if
\begin{equation}\label{ivcmuz}
\mu^{(0)}=0 \qor z^{(0)}=0.
\end{equation}
If $ z^{(0)}=0 $, then $ v^{u(0)}=1 $ and $ v^{r(0)}=0 $. In this case, $ v^a $ coincides with the observer's four-velocity on the worldline, which justifies using the same notation for both.

\section{Kinematics of light cones}\label{kinematics}

This appendix gives the kinematic quantities of section~\ref{imf} in Bondi--Sachs coordinates. The transversal null vector field \eqref{ldef}, the projector onto the surfaces $\mathcal{S}(u,r)$, and the shear scalar are, respectively,
\begin{equation}\label{l}
l^\mu = \left( e^{-2 \beta },\frac{1}{2} e^{2 (\Phi - \beta) },e^{-2 \beta } U,e^{-2 \beta } W\right),
\end{equation}
\[ q^{\mu}{}_{\nu}= \left(
\begin{array}{cccc}
0 & 0 & 0 & 0 \\
0 & 0 & 0 & 0 \\
-U & 0 & 1 & 0 \\
-W & 0 & 0 & 1 \\
\end{array}
\right), \]
\begin{equation}\label{shear2}
\sigma^2\coloneqq\sigma^{a b} \sigma_{a b}=2\left[\cosh ^2(2 \delta)\left(\partial_r \gamma\right)^2+\left(\partial_r \delta\right)^2\right].
\end{equation}

The components of the H\'aj\'i\v{c}ek one-form are
\[ \Omega^H{}_{\theta} =\partial _{\theta } \beta -\frac{1}{2} e^{-2 \beta } r^2 \left(e^{2 \gamma } \cosh (2 \delta ) \partial _r U +\sinh (2 \delta ) \sin (\theta ) \partial _r W \right), \]
\[ \Omega^H{}_{\phi} =  \partial _{\phi } \beta -\frac{1}{2}  e^{-2 \beta } r^2\sin (\theta ) \left( \sinh (2 \delta ) \partial _r U +e^{-2 \gamma }\cosh (2 \delta ) \sin (\theta ) \partial _r W \right), \]
and, with the solution in Appendix~\ref{firstterms}, their lowest-order coefficients are
\begin{equation}\label{Omegatheta}
 \Omega^H{}_{\theta}{}^{(2)} = 2\pi  \partial _{\theta } T_{{rr}}{}^{{(0)}}+ 4 \pi  T_{{r\theta }}{}^{{(1)}} -\frac{1}{2} \left(2 \gamma ^{{(2)}} \cot (\theta )+\partial _{\theta } \gamma ^{{(2)}}+\csc (\theta ) \partial _{\phi } \delta ^{{(2)}}\right),
\end{equation}
\begin{equation}\label{Omegaphi}
\Omega^H{}_{\phi}{}^{(2)} = 2\pi  \partial _{\phi } T_{{rr}}{}^{{(0)}}+4 \pi  T_{{r\phi }}{}^{{(1)}}+\frac{1}{2} \left(\partial _{\phi } \gamma ^{{(2)}}-2 \delta ^{{(2)}} \cos (\theta )-\sin (\theta ) \partial _{\theta } \delta ^{{(2)}}\right).
\end{equation}

The reduced transversal expansion scalar \eqref{xidef} is
\begin{equation}\label{x}
\xi = \frac{2}{r} \left( e^{2( \Phi - \beta) }-1\right)+2 e^{-2 \beta } \left(\partial _{\theta } U+U \cot (\theta )+\partial _{\phi } W\right),
\end{equation}
and, since $\xi^{(0)} = 0$ by the central conditions \eqref{ccg}, its lowest-order coefficient is
\begin{equation}\label{xi1}
\begin{aligned}
\xi^{(1)} = \frac{1}{3}\Big[&16\pi T_{ur}{}^{(0)} - 2\Lambda + \partial_{\theta\theta}\gamma^{(2)} + 3\cot(\theta)\,\partial_\theta\gamma^{(2)} - \frac{\partial_{\phi\phi}\gamma^{(2)}}{\sin^2(\theta)} - 2\gamma^{(2)}\\
&+ \frac{2\,\partial_{\theta\phi}\left(\sin(\theta)\,\delta^{(2)}\right)}{\sin^2(\theta)} - \frac{\partial_\theta\left(\sin(\theta)\,\Omega^H{}_\theta{}^{(2)}\right) + \csc(\theta)\,\partial_\phi\Omega^H{}_\phi{}^{(2)}}{\sin(\theta)}\Big].
\end{aligned}
\end{equation}

From definition \eqref{kappa}, one finds that
\begin{equation}\label{kappadelrbeta}
\kappa=\Gamma^r{}_{rr} = g^{ru}\partial_r g_{ru} = 2 \partial_r \beta.
\end{equation}
Also, by equation \eqref{detg}, $\nabla_a \partial_{(r)}{}^a = 2\partial_r \beta + 2/r$. Since $\nabla_a \partial_{(r)}{}^a = \kappa + \Theta$ \cite[eq. 5.69]{Gourgoulhon2006}, it follows that
\begin{equation}\label{Theta}
\Theta =  \frac{2}{r}.
\end{equation}

Now, consider the relation between the affine parameter $ \lambda' $ and the areal radius $ r $. Since they parametrise the same generators of each light cone, their corresponding null vector fields, $ k^a $ and $ \partial_{(r)}{}^a $, are collinear. Thus, $k^a = \chi\, \partial_{(r)}{}^a$, and the geodesic equation for $k^a$ implies $\partial_r \ln(\chi) = -\kappa = -2\partial_r \beta$ \cite[eq. 2.25]{Gourgoulhon2006}, so that $\chi = \chi_0 e^{-2\beta}$ for some function $\chi_0$ that does not depend on $r$. Since $1 = k^a v_a = \chi e^{2\beta}$ along $r = 0$, we have $\chi_0 = 1$, that is, $\chi = e^{-2\beta}$. Then, $k^\mu = \partial_{(\lambda')}{}^\mu$ implies that \cite{Maedler2016}
\begin{equation}\label{delrlambda}
\partial_r \lambda' = e^{2\beta} \qand \partial_{\lambda'}r = e^{-2\beta} . 
\end{equation}

Another important relation is given by the product
\begin{equation}\label{Ray}
\kappa \Theta = \frac{4\partial_r \beta}{r}=  \sigma^2+8 \pi  {T}_{{rr}},
\end{equation}
where the second equality follows from equations \eqref{beta} and \eqref{shear2}. In fact, the above equation is the Raychaudhuri equation \cite[eq. 6.9]{Gourgoulhon2006}, which takes a simpler form in Bondi--Sachs coordinates, since the expansion scalar becomes a trivial function of the areal radius and $\partial_r \Theta = -\Theta^2/2$. Given the central conditions \eqref{ccg}, we have that
\begin{equation}\label{kT0}
\left(\kappa \Theta \right)^{(0)} = 8 \pi  {T}_{{rr}}{}^{(0)}.
\end{equation}

One can invert equations \eqref{xi1} and \eqref{kT0} to get an expression for $ T_{{ur}}{}^{{(0)}} $ and $ {T}_{{rr}}{}^{(0)} $, respectively. Similarly, equations
\eqref{Omegatheta} and \eqref{Omegaphi} provide expressions for $ T_{{r\theta }}{}^{{(1)}} $ and $ T_{{r\phi}}{}^{{(1)}} $, from which we can calculate $\partial_\theta T_{{r\theta }}{}^{{(1)}} $ and $\partial_\phi T_{{r\phi}}{}^{{(1)}} $. Using these results, it is possible to get equation \eqref{rhokine}.

\section{Complete field equations}\label{appfs}

For the metric \eqref{g}, the hypersurface equations \eqref{her}--\eqref{heu}, the evolution equations \eqref{gammaur} and \eqref{deltaur}, and the matter evolution equations \eqref{meer}--\eqref{meeu} read, in this order,

\begin{equation}\label{beta}
\partial _r \beta =\frac{r}{2}  \left(\cosh ^2(2 \delta ) \left(\partial _r \gamma \right){}^2+\left(\partial _r \delta \right){}^2+4 \pi  {T}_{{rr}}\right),
\end{equation}
\begin{equation}\label{UW1}
\begin{aligned}
&\partial _r \left(e^{-2 \beta } r^4 \left(e^{2 \gamma } \cosh (2 \delta ) \partial _r U +  \sinh (2 \delta )\sin (\theta )\partial _r W \right)\right)\\
&\; = 4 r \partial _{\theta } \beta + r^2 \left(\left(\partial _r \gamma \right) \left(-2 e^{2 \gamma } \csc (\theta ) \left(\sinh (4 \delta ) \partial _{\phi } \gamma +2 \cosh (4 \delta ) \partial _{\phi } \delta \right)+\right .\right .\\
&\; \phantom{{}={}} \; \left .4 \cosh ^2(2 \delta ) \left(\cot (\theta )-\partial _{\theta } \gamma \right)+4 \sinh (4 \delta ) \partial _{\theta } \delta \right)-4 \left(\partial _{\theta } \delta \right) \partial _r \delta +\\
&\; \phantom{{}={}} \; \; e^{2 \gamma } \csc (\theta ) \left(4 \left(\partial _{\phi } \gamma \right) \partial _r \delta -\sinh (4 \delta ) \partial _{{r\phi }} \gamma +2 \partial _{{r\phi }} \delta \right)-2 \partial _{{r\theta }} \beta +\\
&\; \phantom{{}={}} \; \left .2 \cosh ^2(2 \delta ) \partial _{{r\theta }} \gamma -16 \pi  {T}_{{r\theta }}\right),
\end{aligned}
\end{equation}
\begin{equation}\label{UW2}
\begin{aligned}
& \partial _r \left(e^{-2 \beta } r^4 \left(e^{-2 \gamma }\cosh (2 \delta )  \sin (\theta ) \partial _r W +  \sinh (2 \delta ) \partial _r U \right)\right)\\
& \; = e^{-2 \gamma } r^2 \left(4 \left(\partial _r \delta \right) \left(\cot (\theta )-\partial _{\theta } \gamma \right)+2 \left(\partial _r \gamma \right) \left(\sinh (4 \delta ) \left(\cot (\theta )-\partial _{\theta } \gamma \right)+2 \cosh (4 \delta ) \partial _{\theta } \delta \right)+\right .\\
&\; \phantom{{}={}}\;\left . \sinh (4 \delta ) \partial _{{r\theta }} \gamma +2 \partial _{{r\theta }} \delta \right)-2 r \csc (\theta ) \left(r \left(2 \left(\partial _{\phi } \delta \right) \left(\sinh (4 \delta ) \partial _r \gamma +\partial _r \delta \right)+\right .\right .\\
&\; \phantom{{}={}} \left .\left .2 \cosh ^2(2 \delta ) \left(\partial _{\phi } \gamma \right) \partial _r \gamma +\partial _{{r\phi }} \beta +\cosh ^2(2 \delta ) \partial _{{r\phi }} \gamma +8 \pi  {T}_{{r\phi }}\right)-2 \partial _{\phi } \beta \right),
\end{aligned}
\end{equation}
\begin{equation}\label{Phi}
\begin{aligned}
& \partial _r e^{2 \Phi } + e^{2 \Phi } \frac{\left(1-4 \pi  r^2 {T}_{{rr}}\right)}{r} \\
&= r \left(U \left(\partial _{{r\theta }} \gamma +2 \left(\partial _r \gamma \right) \left(\cot (\theta )-\partial _{\theta } \gamma \right)\right)+W \left(-\partial _{{r\phi }} \gamma -2 \left(\partial _r \gamma \right) \partial _{\phi } \gamma \right)\right) \cosh ^2(2 \delta )+\\
&\phantom{{}={}} \frac{1}{4r} e^{-2 (\beta +\gamma )} \cosh (2 \delta ) \left(\left(\sin ^2(\theta ) \left(2 W \left(2 \partial _r W \left(r \left(\partial _r \beta +\partial _r \gamma \right)-2\right)-r \partial _{{rr}} W\right)-\right .\right .\right .\\
&\phantom{{}={}}\left .\left .\left . r \left(\partial _r W\right){}^2\right) -\right .\right .\\
&\phantom{{}={}}  4 e^{2 \gamma } r \left(W \partial _r U +U \partial _r W \right) \left(\partial _r \delta \right) \sin (\theta )+\\
&\phantom{{}={}} \left . e^{4 \gamma } \left(2 U \left(2 \left(\partial _r U\right) \left(r \left(\partial _r \beta -\partial _r \gamma \right)-2\right)-r \partial _{{rr}} U \right)-r \left(\partial _r U\right){}^2\right)\right) r^3+\\
&\phantom{{}={}} 4 e^{4 \beta } \left(-e^{4 \gamma } \left(\left(\partial _{\phi } \beta \right){}^2+2 \left(\partial _{\phi } \gamma \right) \partial _{\phi } \beta +2 \left(\partial _{\phi } \gamma \right){}^2+\right .\right .\\
&\phantom{{}={}}\left .\left . 2 \left(\partial _{\phi } \delta \right){}^2+\partial _{\phi \phi } \beta +\partial _{\phi \phi } \gamma \right) \csc ^2(\theta )+\right .\\
&\phantom{{}={}} 2 e^{2 \gamma } \left(\left(\partial _{\phi } \beta +\partial _{\phi } \gamma \right) \partial _{\theta } \delta +\partial _{\theta \phi } \delta +\left(\cot (\theta )+\partial _{\theta } \beta -\partial _{\theta } \gamma \right) \partial _{\phi } \delta \right) \csc (\theta )-\left(\partial _{\theta } \beta \right){}^2-\\
&\phantom{{}={}} 2 \left(\partial _{\theta } \gamma \right){}^2-2 \left(\partial _{\theta } \delta \right){}^2-\cot (\theta ) \partial _{\theta } \beta -\partial _{\theta \theta } \beta +3 \cot (\theta ) \partial _{\theta } \gamma +2 \left(\partial _{\theta } \beta \right) \partial _{\theta } \gamma +\partial _{\theta \theta } \gamma +\\
&\phantom{{}={}} \left .\left .1\right)\right)-e^{2 \gamma }\cosh (2 \delta ) r U \csc (\theta ) \left(\partial _{{r\phi }} \gamma \right) \sinh (2 \delta ) -2 \partial _{\theta } U -\\
&\phantom{{}={}} \frac{1}{2} r \left(2 e^{2 \beta } \Lambda +\cot (\theta ) \partial _r U +\partial _{{r\theta }} U+\partial _{{r\phi }} W\right)-2 \partial _{\phi } W +2 W \partial _{\phi } \beta +U \left(-2 \cot (\theta )+\right .\\
&\phantom{{}={}} \left .2 \partial _{\theta } \beta +r \left(-\partial _{{r\theta }} \beta +e^{2 \gamma } \csc (\theta ) \left(2 \left(\partial _{\phi } \gamma \right) \partial _r \delta +\partial _{{r\phi }} \delta \right)-2 \left(\partial _r \delta \right) \partial _{\theta } \delta \right)\right)-\\
&\phantom{{}={}} 2 e^{-2 \gamma } r \cosh (4 \delta ) \left(\partial _r \gamma \right) \left(e^{4 \gamma } U \csc ^2(\theta ) \partial _{\phi } \delta -W \partial _{\theta } \delta \right) \sin (\theta )+\\
&\phantom{{}={}} e^{-2 \gamma } r W \left(-e^{2 \gamma } \partial _{{r\phi }} \beta +\left(\partial _{{r\theta }} \delta \right) \sin (\theta )+2 \left(\partial _r \delta \right) \left(\cos (\theta )-e^{2 \gamma } \partial _{\phi } \delta -\left(\partial _{\theta } \gamma \right) \sin (\theta )\right)\right)+\\
&\phantom{{}={}} \frac{1}{2} e^{-2 (\beta +\gamma )} \frac{1}{r} \left(-2 W \left(\partial _r W\right) \left(\partial _r \delta \right) \sin ^2(\theta ) r^4-\right .\\
&\phantom{{}={}} 2 e^{4 \beta } \left(\left(3 \cot (\theta )+2 \partial _{\theta } \beta -4 \partial _{\theta } \gamma \right) \partial _{\theta } \delta +\partial _{\theta \theta } \delta \right)+\\
&\phantom{{}={}} 2 e^{4 \gamma } \left(-U \left(\partial _r U\right) \left(\partial _r \delta \right) r^4-e^{4 \beta } \csc ^2(\theta ) \left(2 \left(\partial _{\phi } \beta +2 \partial _{\phi } \gamma \right) \partial _{\phi } \delta +\partial _{\phi \phi } \delta \right)\right)+\\
&\phantom{{}={}} e^{2 \gamma } \left(\left(-r U \partial _{{rr}} W+\partial _r W\left(2 U \left(r \partial _r \beta-2\right)-r \partial _r U\right)+\right .\right .\\
&\phantom{{}={}}\left .\left . W \left(2 \partial _r U \left(r \partial _r \beta -2\right)-r \partial _{{rr}} U\right)\right) r^3+\right .\\
&\phantom{{}={}}\left .\left . 4 e^{4 \beta } \csc ^2(\theta ) \left(\partial _{\theta \phi } \beta +\left(\partial _{\theta } \beta \right) \partial _{\phi } \beta +2 \left(\partial _{\theta } \delta \right) \partial _{\phi } \delta \right)\right) \sin (\theta )\right) \sinh (2 \delta )+\\
&\phantom{{}={}} \frac{1}{2} r \left(e^{-2 \gamma } W \left(\left(\partial _{{r\theta }} \gamma \right) \sin (\theta )+2 \left(\partial _r \gamma \right) \left(\cos (\theta )-2 e^{2 \gamma } \partial _{\phi } \delta -\left(\partial _{\theta } \gamma \right) \sin (\theta )\right)\right)-\right .\\
&\phantom{{}={}}\left . 2 U \left(\partial _r \gamma \right) \left(e^{2 \gamma } \csc (\theta ) \partial _{\phi } \gamma -2 \partial _{\theta } \delta \right)\right) \sinh (4 \delta )+8 \pi  r {T}_{{ur}},
\end{aligned}
\end{equation}
\begin{equation}\label{esmun1} 
\begin{aligned}
& \partial_r \left( r \partial _{{u}} \gamma  \right) + 2 \tanh (2 \delta ) \left(\left(\partial _r \delta \right) \left(r \partial _u \gamma  \right)+\left(\partial _r \gamma \right) (r \partial _u \delta )\right)\\
&= \frac{1}{8} \frac{1}{r} e^{-2 (\beta +\gamma )} {\sech}(2 \delta ) \left(-8 e^{4 \beta +2 \gamma } \csc (\theta ) \left(\left(\partial _{\theta } \beta \right) \partial _{\phi } \delta -\left(\partial _{\phi } \beta \right) \partial _{\theta } \delta \right)+\right .\\
&\phantom{{}={}} e^{4 \gamma } \left(4 e^{4 \beta } \csc ^2(\theta ) \left(\left(\partial _{\phi } \beta \right){}^2+\partial _{\phi \phi } \beta +4 \pi  {T}_{\phi \phi }\right)-r^4 \left(\partial _r U\right){}^2\right)+r^4 \sin ^2(\theta ) \left(\partial _r W\right){}^2-\\
&\phantom{{}={}} \left .4 e^{4 \beta } \left(\left(\partial _{\theta } \beta \right){}^2-\cot (\theta ) \partial _{\theta } \beta +\partial _{\theta \theta } \beta +4 \pi  {T}_{\theta \theta }\right)\right)+\\
&\phantom{{}={}} \tanh (2 \delta ) \left(\frac{1}{4} e^{-2 \gamma } \left(-8 r e^{2 (\gamma +\Phi )} \left(\partial _r \gamma \right) \partial _r \delta -\sin (\theta ) \left(r \partial _{{r\theta }} W +2 \partial _{\theta } W \right)+\right .\right .\\
&\phantom{{}={}} \left . e^{4 \gamma } \csc (\theta ) \left(r \partial _{{r\phi }} U +2 \partial _{\phi } U \right)\right)+\\
&\phantom{{}={}} r \left(\left(\partial _r \delta \right) \left(U \left(\cot (\theta )-2 \partial _{\theta } \gamma \right)-\partial _{\theta } U -2 W \partial _{\phi } \gamma +\partial _{\phi } W\right)-\right .\\
&\phantom{{}={}} \left .\left .2 \left(\partial _r \gamma \right) \left(U \partial _{\theta } \delta +W \partial _{\phi } \delta \right)\right)\right)+\\
&\phantom{{}={}} \frac{1}{4} \left(-4 e^{2 \Phi } \partial _r \gamma +2 U \left(-2 \partial _{\theta } \gamma +\cot (\theta ) \left(1-r \partial _r \gamma \right)-2 r \partial _{{r\theta }} \gamma \right)+\right .\\
&\phantom{{}={}} r \left(-2 e^{2 \Phi } \left(2 \left(\partial _r \gamma \right) \partial _r \Phi +\partial _{{rr}} \gamma \right)-2 \left(\partial _r \gamma \right) \partial _{\theta } U +\right .\\
&\phantom{{}={}} \left .4 e^{-2 \gamma } \sin (\theta ) \left(\partial _r \delta \right) \left(e^{4 \gamma } \csc ^2(\theta ) \partial _{\phi } U -\partial _{\theta } W\right)-\partial _{{r\theta }} U -4 W \partial _{{r\phi }} \gamma +\partial _{{r\phi }} W\right)+\\
&\phantom{{}={}} \left .r \partial _r U \left(\cot (\theta )-2 \partial _{\theta } \gamma \right)+\left(2-2 r \partial _r \gamma \right) \partial _{\phi } W -2 \partial _{\phi } \gamma \left(r \partial _r W+2 W\right)-2 \partial _{\theta } U\right),
\end{aligned}
 \end{equation}
\begin{equation}\label{esmun2} 
\begin{aligned}
& \partial_r \left( r \partial _{{u}} \delta  \right) - \sinh (4 \delta ) \left(\partial _r \gamma \right) \left( r\partial _u \gamma  \right) \\
&\; = \frac{1}{8} \frac{1}{r} e^{-2 (\beta +\gamma )} \sinh (2 \delta ) \left(4 e^{4 \beta } \left(\left(\partial _{\theta } \beta \right){}^2-\cot (\theta ) \partial _{\theta } \beta +\partial _{\theta \theta } \beta +\right .\right .\\
&\; \phantom{{}={}}\left .\left . e^{4 \gamma } \csc ^2(\theta ) \left(\left(\partial _{\phi } \beta \right){}^2+\partial _{\phi \phi } \beta +4 \pi  {T}_{\phi \phi }\right)+4 \pi  {T}_{\theta \theta }\right)-\right .\\
&\; \phantom{{}={}}\left . r^4 \left(e^{4 \gamma } \left(\partial _r U\right){}^2+\sin ^2(\theta ) \left(\partial _r W\right){}^2\right)\right)+\\
&\; \phantom{{}={}} \frac{1}{4} e^{-2 \beta } \frac{1}{r} \cosh (2 \delta ) \left(r^4 \sin (\theta ) \left(-\partial _r U \right) \partial _r W -\right .\\
&\; \phantom{{}={}}\left . 4 e^{4 \beta } \csc (\theta ) \left(-\left(\partial _{\theta } \beta \right) \partial _{\phi } \gamma +\left(\partial _{\phi } \beta \right) \left(\partial _{\theta } \beta +\partial _{\theta } \gamma -\cot (\theta )\right)+\partial _{\theta \phi } \beta +4 \pi  {T}_{\theta \phi }\right)\right)+\\
&\; \phantom{{}={}} \frac{1}{4} e^{-2 \gamma } \left(-2 e^{2 \gamma } \left(r \left(\left(\partial _r \delta \right) \left(\partial _{\theta } U+U \cot (\theta )+\partial _{\phi } W\right)+\left(\partial _{\phi } \delta \right) \partial _r W +2 U \partial _{{r\theta }} \delta \right)+\right .\right .\\
&\; \phantom{{}={}} \left .2 W \left(\partial _{\phi } \delta +r \partial _{{r\phi }} \delta \right)+\left(\partial _{\theta } \delta \right) \left(r \partial _r U +2 U\right)\right)+\sin (\theta ) \left(-r \partial _{{r\theta }} W -2 \partial _{\theta } W \right)-\\
&\; \phantom{{}={}} \left .2 e^{2 (\gamma +\Phi )} \left(2 \left(\partial _r \delta \right) \left(r \partial _r \Phi +1\right)+r \partial _{{rr}} \delta \right)-e^{4 \gamma } \csc (\theta ) \left(r \partial _{{r\phi }} U +2 \partial _{\phi } U \right)\right)+\\
&\; \phantom{{}={}} \frac{1}{2} r \sinh (4 \delta ) \left(\partial _r \gamma \right) \left(e^{2 \Phi } \partial _r \gamma +U \left(2 \partial _{\theta } \gamma -\cot (\theta )\right)+\partial _{\theta } U+2 W \partial _{\phi } \gamma -\partial _{\phi } W\right)-\\
&\; \phantom{{}={}} e^{-2 \gamma } r \cosh ^2(2 \delta ) \sin (\theta ) \left(\partial _r \gamma \right) \left(e^{4 \gamma } \csc ^2(\theta ) \partial _{\phi } U -\partial _{\theta } W\right),
\end{aligned}
 \end{equation}
\begin{equation}\label{deluTrr}
\begin{aligned}
\partial _u {T}_{{rr}}= &\; e^{2 \beta } \frac{1}{r^3} \sinh (2 \delta ) \left(r \left(-e^{-2 \gamma } \left({T}_{\theta \theta } \partial _r \delta +2 {T}_{{r\theta }} \partial _{\theta } \delta \right)-\right .\right .\\
&\; e^{2 \gamma } \csc ^2(\theta ) \left({T}_{\phi \phi } \partial _r \delta +2 {T}_{{r\phi }} \partial _{\phi } \delta \right)+\\
&\; \left .\left .\csc (\theta ) \left(2 {T}_{{r\theta }} \partial _{\phi } \beta +\partial _{\phi } {T}_{{r\theta }}+2 {T}_{{r\phi }} \partial _{\theta } \beta +\partial _{\theta } {T}_{{r\phi }}\right)\right)-2 \csc (\theta ) {T}_{\theta \phi }\right)+\\
&\; \frac{1}{r^3} e^{2 (\beta -\gamma )} \cosh (2 \delta ) \left(r {T}_{{r\theta }} \left(-2 \partial _{\theta } \beta +2 e^{2 \gamma } \csc (\theta ) \partial _{\phi } \delta +2 \partial _{\theta } \gamma -\cot (\theta )\right)-\right .\\
&\; r \partial _{\theta } {T}_{{r\theta }} -e^{4 \gamma } \csc ^2(\theta ) \left(r \left(2 {T}_{{r\phi }} \left(\partial _{\phi } \beta +\partial _{\phi } \gamma \right)+\partial _{\phi } {T}_{{r\phi }}\right)+{T}_{\phi \phi } \left(r \partial _r \gamma -1\right)\right)+\\
&\; \left .2 e^{2 \gamma } r \csc (\theta ) \left({T}_{\theta \phi } \partial _r \delta +{T}_{{r\phi }} \partial _{\theta } \delta \right)+{T}_{\theta \theta } \left(r \partial _r \gamma +1\right)\right)+\\
&\; \frac{1}{r} \left(2 r \left(\partial _r \beta \right) \left(U {T}_{{r\theta }}+W {T}_{{r\phi }}+{T}_{{ur}}\right)-2 {T}_{{r\theta }} \left(r \partial _r U +U\right)-\right .\\
&\; r \left(U \left(\partial _r {T}_{{r\theta }}+\partial _{\theta } {T}_{{rr}}\right)+e^{2 \Phi } \partial _r {T}_{{rr}} +W \partial _r {T}_{{r\phi }} +2 {T}_{{r\phi }} \partial _r W +\partial _r {T}_{{ur}}\right)+\\
&\; W \left(-r \partial _{\phi } {T}_{{rr}} -2 {T}_{{r\phi }}\right)+\\
&\; \left . {T}_{{rr}} \left(e^{2 \Phi } \left(r \left(\partial _r \beta -3 \partial _r \Phi \right)-2\right)-r \left(\partial _{\theta } U+U \cot (\theta )+\partial _{\phi } W\right)\right)-2 {T}_{{ur}}\right),
\end{aligned}
\end{equation}
\begin{equation}\label{esmun3} 
\begin{aligned}
\partial _u {T}_{{r\theta }}= &\;  \frac{e^{2 \beta }}{r^2} \sinh (2 \delta ) \left(-\csc (\theta ) \left({T}_{\theta \phi } \left(\cot (\theta )-2 \partial _{\theta } \beta \right)-2 {T}_{\theta \theta } \partial _{\phi } \beta -\partial _{\theta } {T}_{\theta \phi }-\partial _{\phi } {T}_{\theta \theta } \right)-\right .\\
&\; \left .3 e^{-2 \gamma } {T}_{\theta \theta } \partial _{\theta } \delta -e^{2 \gamma } \csc ^2(\theta ) \left({T}_{\phi \phi } \partial _{\theta } \delta +2 {T}_{\theta \phi } \partial _{\phi } \delta \right)\right)+\\
&\; \frac{1}{r^2} e^{2 (\beta -\gamma )} \cosh (2 \delta ) \left({T}_{\theta \theta } \left(-2 \partial _{\theta } \beta +3 \partial _{\theta } \gamma -\cot (\theta )\right)+\right .\\
&\; e^{4 \gamma } \csc ^2(\theta ) \left(-2 {T}_{\theta \phi } \left(\partial _{\phi } \beta +\partial _{\phi } \gamma \right)+{T}_{\phi \phi } \left(\cot (\theta )-\partial _{\theta } \gamma \right)-\partial _{\phi } {T}_{\theta \phi } \right)+\\
&\; \left . 2 e^{2 \gamma } \csc (\theta ) \left(2 {T}_{\theta \phi } \partial _{\theta } \delta +{T}_{\theta \theta } \partial _{\phi } \delta \right)-\partial _{\theta } {T}_{\theta \theta }\right)+\\
&\; \frac{1}{r} \left({T}_{{r\theta }} \left(-r \left(U \left(\cot (\theta )-2 \partial _{\theta } \beta \right)+2 \partial _{\theta } U +\partial _{\phi } W\right)-2 e^{2 \Phi } \left(r \partial _r \Phi +1\right)\right)+\right .\\
&\;  r \left(W \left(-\partial _r {T}_{\theta \phi }-\partial _{\phi } {T}_{{r\theta }} +2 {T}_{{r\phi }} \partial _{\theta } \beta \right)+U \left(-\partial _r {T}_{\theta \theta }-\partial _{\theta } {T}_{{r\theta }} \right)+\right .\\
&\; \left .e^{2 \Phi } \left({T}_{{rr}} \left(\partial _{\theta } \beta -\partial _{\theta } \Phi \right)-\partial _r {T}_{{r\theta }}\right)-\partial _r {T}_{{u\theta }}-{T}_{\theta \phi } \partial _r W-{T}_{{r\phi }} \partial _{\theta } W+2 {T}_{{ur}} \partial _{\theta } \beta \right)+\\
&\; \left .{T}_{\theta \theta } \left(-r \partial _r U -2 U\right)-2 \left({T}_{{u\theta }}+W {T}_{\theta \phi }\right)\right),
\end{aligned}
 \end{equation}
\begin{equation}\label{esmun4} 
\begin{aligned}
\partial _u {T}_{{r\phi }}= &\; e^{2 \beta } \frac{1}{r^2} \sinh (2 \delta ) \left(2 {T}_{\theta \phi } \left(\csc (\theta ) \partial _{\phi } \beta -e^{-2 \gamma } \partial _{\theta } \delta \right)-\right .\\
&\; \csc (\theta ) \left({T}_{\phi \phi } \left(3 e^{2 \gamma } \csc (\theta ) \partial _{\phi } \delta -2 \partial _{\theta } \beta \right)-\partial _{\theta } {T}_{\phi \phi }\right)-e^{-2 \gamma } {T}_{\theta \theta } \partial _{\phi } \delta +\\
&\; \left .\csc (\theta ) \partial _{\phi } {T}_{\theta \phi } \right)+\frac{1}{r^2} e^{2 (\beta -\gamma )} \cosh (2 \delta ) \left({T}_{\theta \phi } \left(-2 \partial _{\theta } \beta +2 \partial _{\theta } \gamma -\cot (\theta )\right)-\right .\\
&\; e^{4 \gamma } \csc ^2(\theta ) \left({T}_{\phi \phi } \left(2 \partial _{\phi } \beta +3 \partial _{\phi } \gamma \right)+\partial _{\phi } {T}_{\phi \phi }\right)+2 e^{2 \gamma } \csc (\theta ) \left({T}_{\phi \phi } \partial _{\theta } \delta +2 {T}_{\theta \phi } \partial _{\phi } \delta \right)+\\
&\; \left . {T}_{\theta \theta } \partial _{\phi } \gamma -\partial _{\theta } {T}_{\theta \phi }\right)+\frac{1}{r} \left(r \left(e^{2 \Phi } \left({T}_{{rr}} \left(\partial _{\phi } \beta -\partial _{\phi } \Phi \right)-\partial _r {T}_{{r\phi }}\right)+\right .\right .\\
&\; U \left(-\partial _r {T}_{\theta \phi }-\partial _{\theta } {T}_{{r\phi }} \right)+W \left(-\partial _r {T}_{\phi \phi }-\partial _{\phi } {T}_{{r\phi }} \right)-\partial _r {T}_{{u\phi }} -{T}_{\phi \phi } \partial _r W +\\
&\; \left .{T}_{{r\theta }} \left(2 U \partial _{\phi } \beta -\partial _{\phi } U\right)+2 {T}_{{ur}} \partial _{\phi } \beta \right)+\\
&\; {T}_{{r\phi }} \left(-r \left(U \cot (\theta )+\partial _{\theta } U-2 W \partial _{\phi } \beta +2 \partial _{\phi } W \right)-2 e^{2 \Phi } \left(r \partial _r \Phi +1\right)\right)+\\
&\; \left . {T}_{\theta \phi } \left(-r \partial _r U -2 U\right)-2 \left({T}_{{u\phi }}+W {T}_{\phi \phi }\right)\right),
\end{aligned}
 \end{equation}
\begin{equation}\label{deluTur}
\begin{aligned}
\partial _u {T}_{{ur}}=&\; \frac{e^{2 (\beta -\gamma )} }{r^2} \cosh (2 \delta ) \left({T}_{\theta \theta } \partial _u \gamma -e^{4 \gamma } \csc ^2(\theta ) \left({T}_{\phi \phi } \partial _u \gamma +2 {T}_{{u\phi }} \left(\partial _{\phi } \beta +\partial _{\phi } \gamma \right)+\partial _{\phi } {T}_{{u\phi }}\right)+\right .\\
&\; 2 e^{2 \gamma } \csc (\theta ) \left({T}_{\theta \phi } \partial _u \delta +{T}_{{u\phi }} \partial _{\theta } \delta \right)+\\
&\;\left . {T}_{{u\theta }} \left(-2 \partial _{\theta } \beta +2 e^{2 \gamma } \csc (\theta ) \partial _{\phi } \delta +2 \partial _{\theta } \gamma -\cot (\theta )\right)- \partial _{\theta } {T}_{{u\theta }}\right)+\\
&\; e^{2 \beta } \frac{1}{r^2} \sinh (2 \delta ) \left(-e^{-2 \gamma } \left({T}_{\theta \theta } \partial _u \delta +2 {T}_{{u\theta }} \partial _{\theta } \delta \right)-\right .\\
&\; e^{2 \gamma } \csc ^2(\theta ) \left({T}_{\phi \phi } \partial _u \delta +2 {T}_{{u\phi }} \partial _{\phi } \delta \right)+\\
&\;\left . \csc (\theta ) \left(2 {T}_{{u\theta }} \partial _{\phi } \beta +\partial _{\phi } {T}_{{u\theta }}+2 {T}_{{u\phi }} \partial _{\theta } \beta +\partial _{\theta } {T}_{{u\phi }}\right)\right)+\\
&\; \frac{1}{r} \left(r \left(U \left(-\partial _r {T}_{{u\theta }}+2 {T}_{{r\theta }} \partial _u \beta -\partial _{\theta } {T}_{{ur}} \right)-e^{2 \Phi } \left(\partial _r {T}_{{ur}}-{T}_{{rr}} \partial _u \beta \right)+\right .\right .\\
&\; W \left(-\partial _r {T}_{{u\phi }}+2 {T}_{{r\phi }} \partial _u \beta -\partial _{\phi } {T}_{{ur}} \right)-\partial _r {T}_{{uu}} -{T}_{{u\phi }} \partial _r W -{T}_{{r\theta }} \partial _u U -\\
&\;\left . {T}_{{r\phi }} \partial _u W\right)+{T}_{{ur}} \left(-r \left(-2 \partial _u \beta +\partial _{\theta } U+U \cot (\theta )+\partial _{\phi } W\right)-2 e^{2 \Phi } \left(r \partial _r \Phi +1\right)\right)+\\
&\; \left . {T}_{{u\theta }} \left(-r \partial _r U -2 U\right)-2 \left({T}_{{uu}}+W {T}_{{u\phi }}\right)\right)-\frac{1}{2} {T}_{{rr}} \partial _u e^{2 \Phi } .
\end{aligned}
\end{equation}

\section{First terms of the solution}\label{firstterms}

With the expansion \eqref{f} and the central conditions \eqref{ccg} and \eqref{centralT}, the hierarchical scheme gives the following coefficients, to the first order not fixed by the central conditions.

\begin{equation}\label{beta2}
\beta^{(2)}=2 \pi  T_{{rr}}{}^{{(0)}},
\end{equation}
\begin{equation}\label{U1}
U^{(1)}=-4 \pi  T_{{r\theta }}{}^{{(1)}}+\frac{1}{2} \left(\partial _{\theta } \gamma ^{{(2)}}+\csc (\theta ) \partial _{\phi } \delta ^{{(2)}} \right)+\gamma ^{{(2)}} \cot (\theta ),
\end{equation}
\begin{equation}\label{W1}
W^{(1)}=\csc ^2(\theta ) \left(-4 \pi  T_{{r\phi }}{}^{{(1)}}+\frac{1}{2} \left(\sin (\theta ) \partial _{\theta } \delta ^{{(2)}} -\partial _{\phi } \gamma ^{{(2)}}\right)+\delta ^{{(2)}} \cos (\theta )\right),
\end{equation}
\begin{equation}\label{Phi2}
\begin{aligned}
\Phi^{(2)}& =\frac{1}{3} \left(\pi  \left(\cot (\theta ) \left(10 T_{{r\theta }}{}^{{(1)}}-\partial _{\theta } T_{{rr}}{}^{{(0)}}\right)+\csc ^2(\theta ) \left(10\partial _{\phi } T_{{r\phi }}{}^{{(1)}}-\partial _{\phi \phi } T_{{rr}}{}^{{(0)}}\right)- \right .\right .\\
&\phantom{{}={}}\; \left .\left . \partial _{\theta \theta } T_{{rr}}{}^{{(0)}}+6 T_{{rr}}{}^{{(0)}}+8 T_{{ur}}{}^{{(0)}}+10 \partial _{\theta } T_{{r\theta }}{}^{{(1)}} \right)-\Lambda \right)+\\
&\phantom{{}={}}\; \frac{1}{2} \left(\gamma ^{{(2)}}+\csc (\theta ) \left(-\cot (\theta ) \partial _{\phi } \delta ^{{(2)}} -\partial _{\theta \phi } \delta ^{{(2)}} \right)\right)+\\
&\phantom{{}={}}\; \frac{1}{4} \left(\csc ^2(\theta ) \partial _{\phi \phi } \gamma ^{{(2)}} -3 \cot (\theta ) \partial _{\theta } \gamma ^{{(2)}} -\partial _{\theta \theta } \gamma ^{{(2)}}\right)-2 ( \Phi ^{{(1)}} )^2,
\end{aligned}
\end{equation}
\begin{equation}\label{gamma2u}
\begin{aligned}
\partial_{u}\gamma^{(2)}&=\frac{1}{45} \left(-15 \gamma ^{{(2)}} \left(\partial _{\theta } U^{{(0)}}+U^{{(0)}} \cot (\theta )+\partial _{\phi } W^{{(0)}}+8 \Phi ^{{(1)}}\right)-45 U^{{(0)}} \partial _{\theta } \gamma ^{{(2)}} +\right .\\
&\quad 45 \delta ^{{(2)}} \csc (\theta ) \partial _{\phi } U^{{(0)}} -45 W^{{(0)}} \partial _{\phi } \gamma ^{{(2)}} -45 \delta ^{{(2)}} \sin (\theta ) \partial _{\theta } W^{{(0)}} +\\
&\quad 6 \pi  \csc ^2(\theta ) \partial _{\phi \phi } T_{{rr}}{}^{{(1)}} +6 \pi  \cot (\theta ) \partial _{\theta } T_{{rr}}{}^{{(1)}} -6 \pi  \partial _{\theta \theta } T_{{rr}}{}^{{(1)}} -24 \pi  T_{{r\theta }}{}^{{(2)}} \cot (\theta )+\\
&\quad 24 \pi  \partial _{\theta } T_{{r\theta }}{}^{{(2)}} -24 \pi  \csc ^2(\theta ) \partial _{\phi } T_{{r\phi }}{}^{{(2)}} -3 \csc ^2(\theta ) \partial _{\phi \phi } \gamma ^{{(3)}} +\\
&\quad 12 \gamma ^{{(3)}} \left(\csc ^2(\theta )-3\right)-3 \cot (\theta ) \partial _{\theta } \gamma ^{{(3)}} -3 \partial _{\theta \theta } \gamma ^{{(3)}} +12 \cot (\theta ) \csc (\theta ) \partial _{\phi } \delta ^{{(3)}} -\\
&\quad \left . 10 \pi  T_{\theta \theta }{}^{{(3)}}+10 \pi  T_{\phi \phi }{}^{{(3)}} \csc ^2(\theta )\right),
\end{aligned}
\end{equation}
\begin{equation}\label{delta2u}
\begin{aligned}
\partial_{u}\delta^{(2)}=\;& 
\frac{1}{45} \left(-15 \delta ^{{(2)}} \left(\partial _{\theta } U^{{(0)}}+U^{{(0)}} \cot (\theta )+\partial _{\phi } W^{{(0)}}+8 \Phi ^{{(1)}}\right)-45 \gamma ^{{(2)}} \csc (\theta ) \partial _{\phi } U^{{(0)}} -\right .\\
& 45 U^{{(0)}} \partial _{\theta } \delta ^{{(2)}} +45 \gamma ^{{(2)}} \sin (\theta ) \partial _{\theta } W^{{(0)}} -45 W^{{(0)}} \partial _{\phi } \delta ^{{(2)}} -\\
& 12 \pi  \csc (\theta ) \partial _{\theta \phi } T_{{rr}}{}^{{(1)}} +12 \pi  \cot (\theta ) \csc (\theta ) \partial _{\phi } T_{{rr}}{}^{{(1)}} +24 \pi  \csc (\theta ) \partial _{\phi } T_{{r\theta }}{}^{{(2)}} -\\
& 48 \pi  T_{{r\phi }}{}^{{(2)}} \cot (\theta ) \csc (\theta )+24 \pi  \csc (\theta ) \partial _{\theta } T_{{r\phi }}{}^{{(2)}} -12 \cot (\theta ) \csc (\theta ) \partial _{\phi } \gamma ^{{(3)}} -\\
& 3 \csc ^2(\theta ) \partial _{\phi \phi } \delta ^{{(3)}} +12 \delta ^{{(3)}} \left(\csc ^2(\theta )-3\right)-3 \cot (\theta ) \partial _{\theta } \delta ^{{(3)}} -3 \partial _{\theta \theta } \delta ^{{(3)}} -\\
&\left . 20 \pi  T_{\theta \phi }{}^{{(3)}} \csc (\theta )\right),
\end{aligned}
\end{equation}
\begin{equation}\label{Trru}
\begin{aligned}
\partial_{u}T_{rr}{}^{(0)}= \;&
-3 U^{{(0)}} T_{{r\theta }}{}^{{(1)}}-W^{{(0)}} \left(\partial _{\phi } T_{{rr}}{}^{{(0)}}+3 T_{{r\phi }}{}^{{(1)}}\right)-\\
& \left(T_{{rr}}{}^{{(0)}} \left(\partial _{\theta } U^{{(0)}}+U^{{(0)}} \cot (\theta )+\partial _{\phi } W^{{(0)}}+7 \Phi ^{{(1)}}\right)\right)-U^{{(0)}} \partial _{\theta } T_{{rr}}{}^{{(0)}} -\\
& 3 T_{{rr}}{}^{{(1)}}-3 T_{{ur}}{}^{{(1)}}+\frac{1}{2} \left(T_{{r\theta }}{}^{{(2)}} (-\cot (\theta ))-\partial _{\theta } T_{{r\theta }}{}^{{(2)}}-\csc ^2(\theta ) \partial _{\phi } T_{{r\phi }}{}^{{(2)}} \right)+\\
& \frac{1}{6} \left(T_{\theta \theta }{}^{{(3)}}+T_{\phi \phi }{}^{{(3)}} \csc ^2(\theta )\right),
\end{aligned}
\end{equation}
\begin{equation}\label{Trthetau}
\begin{aligned}
\partial_{u}T_{r\theta}{}^{(1)}=\;& 
-2 T_{{r\theta }}{}^{{(1)}} \left(\partial _{\theta } U^{{(0)}}+4 \Phi ^{{(1)}}\right)-6 U^{{(0)}} \partial _{\theta } T_{{r\theta }}{}^{{(1)}} -4 W^{{(0)}} \partial _{\phi } T_{{r\theta }}{}^{{(1)}} -\\
& T_{{r\theta }}{}^{{(1)}} \partial _{\phi } W^{{(0)}} +T_{{rr}}{}^{{(0)}} \left(-\partial _{\theta } \Phi ^{{(1)}} \right)+U^{{(0)}} \csc ^2(\theta ) \partial _{\phi } T_{{r\phi }}{}^{{(1)}} -\\
& 3 W^{{(0)}} \partial _{\theta } T_{{r\phi }}{}^{{(1)}} +T_{{r\phi }}{}^{{(1)}} \left(6 W^{{(0)}} \cot (\theta )-\partial _{\theta } W^{{(0)}}\right)+\\
& \frac{3}{8 \pi } \left(U^{{(0)}} \left(\csc ^2(\theta ) \partial _{\phi \phi } \gamma ^{{(2)}}+\gamma ^{{(2)}} \left(-\cot ^2(\theta )-3 \csc ^2(\theta )+5\right)+\cot (\theta ) \partial _{\theta } \gamma ^{{(2)}}+\right .\right .\\
&\left . \partial _{\theta \theta } \gamma ^{{(2)}}-4 \cot (\theta ) \csc (\theta ) \partial _{\phi } \delta ^{{(2)}}\right)+W^{{(0)}} \left(4 \cot (\theta ) \partial _{\phi } \gamma ^{{(2)}}+\sin (\theta ) \partial _{\theta \theta } \delta ^{{(2)}}+\right .\\
&\left .\left . \csc (\theta ) \partial _{\phi \phi } \delta ^{{(2)}} -\left(\delta ^{{(2)}} (3 \cos (2 \theta )+1) \csc (\theta )\right)+\cos (\theta ) \partial _{\theta } \delta ^{{(2)}} \right)\right)-\\
& 4 U^{{(0)}} \left(T_{{rr}}{}^{{(0)}}+T_{{ur}}{}^{{(0)}}\right)-\frac{1}{2} U^{{(0)}} \csc ^2(\theta ) \partial _{\phi \phi } T_{{rr}}{}^{{(0)}} -\frac{1}{2} U^{{(0)}} \cot (\theta ) \partial _{\theta } T_{{rr}}{}^{{(0)}}+\\
& \frac{1}{2} U^{{(0)}} \partial _{\theta \theta } T_{{rr}}{}^{{(0)}}-W^{{(0)}} \cot (\theta ) \partial _{\phi } T_{{rr}}{}^{{(0)}}+W^{{(0)}} \partial _{\theta \phi } T_{{rr}}{}^{{(0)}}-2 T_{{r\theta }}{}^{{(2)}}-2 T_{{u\theta }}{}^{{(2)}}+\\
& \frac{1}{6} \left(T_{\theta \theta }{}^{{(3)}} (-\cot (\theta ))-\partial _{\theta } T_{\theta \theta }{}^{{(3)}}-\csc ^2(\theta ) \partial _{\phi } T_{\theta \phi }{}^{{(3)}} +T_{\phi \phi }{}^{{(3)}} \cot (\theta ) \csc ^2(\theta )\right),
\end{aligned}
\end{equation}
\begin{equation}\label{Trphiu}
\begin{aligned}
\partial_{u}T_{r\phi}{}^{(1)}=\;& 
T_{{r\theta }}{}^{{(1)}} \left(-\partial _{\phi } U^{{(0)}}-\frac{5}{2} W^{{(0)}} \sin (2 \theta )\right)-3 U^{{(0)}} \partial _{\phi } T_{{r\theta }}{}^{{(1)}} +\\
& W^{{(0)}} \sin ^2(\theta ) \partial _{\theta } T_{{r\theta }}{}^{{(1)}} +T_{{rr}}{}^{{(0)}} \left(-\partial _{\phi } \Phi ^{{(1)}} \right)-4 U^{{(0)}} \partial _{\theta } T_{{r\phi }}{}^{{(1)}} +\\
& T_{{r\phi }}{}^{{(1)}} \left(-\partial _{\theta } U^{{(0)}}+5 U^{{(0)}} \cot (\theta )-2 \partial _{\phi } W^{{(0)}} -8 \Phi ^{{(1)}}\right)-6 W^{{(0)}} \partial _{\phi } T_{{r\phi }}{}^{{(1)}} +\\
& \frac{1}{16 \pi }\left(6 U^{{(0)}} \left(4 \cot (\theta ) \partial _{\phi } \gamma ^{{(2)}} +\sin (\theta ) \partial _{\theta \theta } \delta ^{{(2)}} +\csc (\theta ) \partial _{\phi \phi } \delta ^{{(2)}} -\right .\right .\\
& \left .\left(\delta ^{{(2)}} (3 \cos (2 \theta )+1) \csc (\theta )\right)+\cos (\theta ) \partial _{\theta } \delta ^{{(2)}} \right)+3 W^{{(0)}} \left(-2 \sin ^2(\theta ) \partial _{\theta \theta } \gamma ^{{(2)}} +\right .\\
& \left .\left .\gamma ^{{(2)}} (6 \cos (2 \theta )+2)-\sin (2 \theta ) \partial _{\theta } \gamma ^{{(2)}} -2 \partial _{\phi \phi } \gamma ^{{(2)}} +8 \cos (\theta ) \partial _{\phi } \delta ^{{(2)}} \right)\right)-\\
& 4 W^{{(0)}} \sin ^2(\theta ) \left(T_{{rr}}{}^{{(0)}}+T_{{ur}}{}^{{(0)}}\right)-U^{{(0)}} \cot (\theta ) \partial _{\phi } T_{{rr}}{}^{{(0)}} +U^{{(0)}} \partial _{\theta \phi } T_{{rr}}{}^{{(0)}} -\\
& \frac{1}{2} W^{{(0)}} \sin ^2(\theta ) \partial _{\theta \theta } T_{{rr}}{}^{{(0)}} +\frac{1}{4} W^{{(0)}} \sin (2 \theta ) \partial _{\theta } T_{{rr}}{}^{{(0)}} +\frac{1}{2} W^{{(0)}} \partial _{\phi \phi } T_{{rr}}{}^{{(0)}} -\\
& 2 T_{{r\phi }}{}^{{(2)}}-2 T_{{u\phi }}{}^{{(2)}}+\frac{1}{6} \left(T_{\theta \phi }{}^{{(3)}} (-\cot (\theta ))-\partial _{\theta } T_{\theta \phi }{}^{{(3)}}-\csc ^2(\theta ) \partial _{\phi } T_{\phi \phi }{}^{{(3)}} \right),
\end{aligned}
\end{equation}
\begin{equation}\label{Truu}
\begin{aligned}
\partial_{u}T_{ru}{}^{(0)}=\;& 
\frac{1}{8} \left(6 U^{{(0)}} \csc ^2(\theta ) \partial _{\phi \phi } T_{{r\theta }}{}^{{(1)}} -2 U^{{(0)}} T_{{r\theta }}{}^{{(1)}} (9 \cos (2 \theta )-4) \csc ^2(\theta )+\right .\\
& 10 U^{{(0)}} \cot (\theta ) \partial _{\theta } T_{{r\theta }}{}^{{(1)}} +10 U^{{(0)}} \partial _{\theta \theta } T_{{r\theta }}{}^{{(1)}} +16 W^{{(0)}} \cot (\theta ) \partial _{\phi } T_{{r\theta }}{}^{{(1)}} +\\
& 4 W^{{(0)}} \partial _{\theta \phi } T_{{r\theta }}{}^{{(1)}}+4 U^{{(0)}} \csc ^2(\theta ) \partial _{\theta \phi } T_{{r\phi }}{}^{{(1)}}-20 U^{{(0)}} \cot (\theta ) \csc ^2(\theta ) \partial _{\phi } T_{{r\phi }}{}^{{(1)}}+\\
& 10 W^{{(0)}} \csc ^2(\theta ) \partial _{\phi \phi } T_{{r\phi }}{}^{{(1)}} -6 W^{{(0)}} \cot (\theta ) \partial _{\theta } T_{{r\phi }}{}^{{(1)}} +6 W^{{(0)}} \partial _{\theta \theta } T_{{r\phi }}{}^{{(1)}} +\\
& 36 W^{{(0)}} T_{{r\phi }}{}^{{(1)}}-48 T_{{ur}}{}^{{(0)}} \Phi ^{{(1)}}-U^{{(0)}} \cot (\theta ) \partial _{\theta \theta } T_{{rr}}{}^{{(0)}} -\\
& U^{{(0)}} \csc ^2(\theta ) \partial _{\theta \phi \phi } T_{{rr}}{}^{{(0)}} +2 U^{{(0)}} \cot (\theta ) \csc ^2(\theta ) \partial _{\phi \phi } T_{{rr}}{}^{{(0)}} +\\
& U^{{(0)}} \left(2 \cot ^2(\theta )-\csc ^2(\theta )+8\right) \partial _{\theta } T_{{rr}}{}^{{(0)}} -U^{{(0)}} \partial _{\theta \theta \theta } T_{{rr}}{}^{{(0)}} -\\
& W^{{(0)}} \cot (\theta ) \partial _{\theta \phi } T_{{rr}}{}^{{(0)}} -W^{{(0)}} \csc ^2(\theta ) \partial _{\phi \phi \phi } T_{{rr}}{}^{{(0)}} -W^{{(0)}} \partial _{\theta \theta \phi } T_{{rr}}{}^{{(0)}} +\\
& 6 W^{{(0)}} \partial _{\phi } T_{{rr}}{}^{{(0)}} -8 T_{{ur}}{}^{{(0)}} U^{{(0)}} \cot (\theta )-8 T_{{ur}}{}^{{(0)}} \partial _{\theta } U^{{(0)}} -8 T_{{ur}}{}^{{(0)}} \partial _{\phi } W^{{(0)}} -\\
&\left . 24 T_{{ur}}{}^{{(1)}}-24 T_{{uu}}{}^{{(1)}}-4 T_{{u\theta }}{}^{{(2)}} \cot (\theta )-4 \partial _{\theta } T_{{u\theta }}{}^{{(2)}} -4 \csc ^2(\theta ) \partial _{\phi } T_{{u\phi }}{}^{{(2)}} \right)+\\
& \frac{3}{32 \pi } \left(U^{{(0)}} \left(-3 \cot (\theta ) \partial _{\theta \theta } \gamma ^{{(2)}}-\csc ^2(\theta ) \partial _{\theta \phi \phi } \gamma ^{{(2)}}-4 \cot (\theta ) \csc ^2(\theta ) \partial _{\phi \phi } \gamma ^{{(2)}}-\right .\right .\\
& 12 \gamma ^{{(2)}} \cot (\theta )-4 \partial _{\theta } \gamma ^{{(2)}} +3 \csc ^2(\theta ) \partial _{\theta } \gamma ^{{(2)}} -\partial _{\theta \theta \theta } \gamma ^{{(2)}}-\csc (\theta ) \partial _{\theta \theta \phi } \delta ^{{(2)}} +\\
& \left .3 \cot (\theta ) \csc (\theta ) \partial _{\theta \phi } \delta ^{{(2)}}+(5 \cos (2 \theta )-1) \csc ^3(\theta ) \partial _{\phi } \delta ^{{(2)}}-\csc ^3(\theta ) \partial _{\phi \phi \phi } \delta ^{{(2)}}\right)+\\
& W^{{(0)}} \left(-3 \cot (\theta ) \partial _{\theta \phi } \gamma ^{{(2)}} -(5 \cos (2 \theta )-1) \csc ^2(\theta ) \partial _{\phi } \gamma ^{{(2)}} +\right .\\
& \csc ^2(\theta ) \partial _{\phi \phi \phi } \gamma ^{{(2)}} +\partial _{\theta \theta \phi } \gamma ^{{(2)}}-3 \cos (\theta ) \partial _{\theta \theta } \delta ^{{(2)}} -\sin (\theta ) \partial _{\theta \theta \theta } \delta ^{{(2)}} -\\
& \csc (\theta ) \partial _{\theta \phi \phi } \delta ^{{(2)}} -4 \cot (\theta ) \csc (\theta ) \partial _{\phi \phi } \delta ^{{(2)}} -12 \delta ^{{(2)}} \cos (\theta )+\csc (\theta ) \partial _{\theta } \delta ^{{(2)}} +\\
&\left .\left . 2 \cos (2 \theta ) \csc (\theta ) \partial _{\theta } \delta ^{{(2)}} \right)\right).
\end{aligned}
\end{equation}

The leading-order terms in $r$ of equations \eqref{gammaur} and \eqref{deltaur} and of the matter evolution equations \eqref{meer}--\eqref{meeu} impose the following initial-value constraints on the coefficients:
\begin{equation}\label{Tthetatheta2}
\begin{aligned}
T_{\theta\theta}{}^{(2)}=\;&
\frac{1}{4} \left(\csc ^2(\theta ) \partial _{\phi \phi } T_{{rr}}{}^{{(0)}} +\cot (\theta ) \partial _{\theta } T_{{rr}}{}^{{(0)}} -\partial _{\theta \theta } T_{{rr}}{}^{{(0)}}+8 T_{{rr}}{}^{{(0)}}+8 T_{{ur}}{}^{{(0)}}-\right .\\
&\left . 2 T_{{r\theta }}{}^{{(1)}} \cot (\theta )+10 \partial _{\theta } T_{{r\theta }}{}^{{(1)}}-2 \csc ^2(\theta ) \partial _{\phi } T_{{r\phi }}{}^{{(1)}}\right)+\frac{3}{16 \pi } \left(-\csc ^2(\theta ) \partial _{\phi \phi } \gamma ^{{(2)}}+\right .\\
& \left .\gamma ^{{(2)}} (3 \cos (2 \theta )+1) \csc ^2(\theta )-\cot (\theta ) \partial _{\theta } \gamma ^{{(2)}} -\partial _{\theta \theta } \gamma ^{{(2)}}+4 \cot (\theta ) \csc (\theta ) \partial _{\phi } \delta ^{{(2)}} \right),
\end{aligned}
\end{equation} 
\begin{equation}\label{Tphiphi2}
\begin{aligned}
T_{\phi\phi}{}^{(2)}=\;&
\frac{1}{8} \left(2 \sin ^2(\theta ) \partial _{\theta \theta } T_{{rr}}{}^{{(0)}} -\sin (2 \theta ) \partial _{\theta } T_{{rr}}{}^{{(0)}} +16 \sin ^2(\theta ) \left(T_{{rr}}{}^{{(0)}}+T_{{ur}}{}^{{(0)}}\right)-\right .\\
&\left . 2 \partial _{\phi \phi } T_{{rr}}{}^{{(0)}} +10 T_{{r\theta }}{}^{{(1)}} \sin (2 \theta )-4 \sin ^2(\theta ) \partial _{\theta } T_{{r\theta }}{}^{{(1)}} +20 \partial _{\phi } T_{{r\phi }}{}^{{(1)}} \right)-\\
& \frac{3}{32 \pi } \left(-2 \sin ^2(\theta ) \partial _{\theta \theta } \gamma ^{{(2)}} +\gamma ^{{(2)}} (6 \cos (2 \theta )+2)-\sin (2 \theta ) \partial _{\theta } \gamma ^{{(2)}} -2 \partial _{\phi \phi } \gamma ^{{(2)}} +\right .\\
&\left .8 \cos (\theta ) \partial _{\phi } \delta ^{{(2)}} \right),
\end{aligned}
\end{equation}
\begin{equation}\label{Tthetaphi2}
\begin{aligned}
T_{\theta\phi}{}^{(2)}=\;&
\frac{1}{2} \left(\cot (\theta ) \partial _{\phi } T_{{rr}}{}^{{(0)}} -\partial _{\theta \phi } T_{{rr}}{}^{{(0)}}+3 \partial _{\phi } T_{{r\theta }}{}^{{(1)}} -6 T_{{r\phi }}{}^{{(1)}} \cot (\theta )+3 \partial _{\theta } T_{{r\phi }}{}^{{(1)}} \right)+\\
& \frac{3}{16 \pi } \left(-4 \cot (\theta ) \partial _{\phi } \gamma ^{{(2)}} -\sin (\theta ) \partial _{\theta \theta } \delta ^{{(2)}} -\csc (\theta ) \partial _{\phi \phi } \delta ^{{(2)}} +\right .\\
& \left .\delta ^{{(2)}} (3 \cos (2 \theta )+1) \csc (\theta )-\cos (\theta ) \partial _{\theta } \delta ^{{(2)}} \right),
\end{aligned}
\end{equation}
\begin{equation}\label{Tutheta1}
\begin{aligned}
T_{u\theta}{}^{(1)}=\;&
\frac{1}{24} \left(\cot (\theta ) \partial _{\theta \theta } T_{{rr}}{}^{{(0)}} +\csc ^2(\theta ) \partial _{\theta \phi \phi } T_{{rr}}{}^{{(0)}} -2 \cot (\theta ) \csc ^2(\theta ) \partial _{\phi \phi } T_{{rr}}{}^{{(0)}} +\right .\\
& (3 \cos (2 \theta )-4) \csc ^2(\theta ) \partial _{\theta } T_{{rr}}{}^{{(0)}} +\partial _{\theta \theta \theta } T_{{rr}}{}^{{(0)}}-8 \partial _{\theta } T_{{ur}}{}^{{(0)}} -\\
& 6 \csc ^2(\theta ) \partial _{\phi \phi } T_{{r\theta }}{}^{{(1)}} +2 T_{{r\theta }}{}^{{(1)}} (9 \cos (2 \theta )-4) \csc ^2(\theta )-10 \cot (\theta ) \partial _{\theta } T_{{r\theta }}{}^{{(1)}} -\\
& \left .10 \partial _{\theta \theta } T_{{r\theta }}{}^{{(1)}} -4 \csc ^2(\theta ) \partial _{\theta \phi } T_{{r\phi }}{}^{{(1)}} +20 \cot (\theta ) \csc ^2(\theta ) \partial _{\phi } T_{{r\phi }}{}^{{(1)}} \right)+\\
& \frac{1}{32 \pi }\left(3 \cot (\theta ) \partial _{\theta \theta } \gamma ^{{(2)}} +\csc ^2(\theta ) \partial _{\theta \phi \phi } \gamma ^{{(2)}} +4 \cot (\theta ) \csc ^2(\theta ) \partial _{\phi \phi } \gamma ^{{(2)}} +\right .\\
& 12 \gamma ^{{(2)}} \cot (\theta )+\left(4-3 \csc ^2(\theta )\right) \partial _{\theta } \gamma ^{{(2)}} +\partial _{\theta \theta \theta } \gamma ^{{(2)}}+\csc (\theta ) \partial _{\theta \theta \phi } \delta ^{{(2)}} -\\
&\left . 3 \cot (\theta ) \csc (\theta ) \partial _{\theta \phi } \delta ^{{(2)}} +(1-5 \cos (2 \theta )) \csc ^3(\theta ) \partial _{\phi } \delta ^{{(2)}} +\csc ^3(\theta ) \partial _{\phi \phi \phi } \delta ^{{(2)}} \right),
\end{aligned}
\end{equation}
\begin{equation}\label{Tuphi1}
\begin{aligned}
T_{u\phi}{}^{(1)}=\;&
\frac{1}{24} \left(\cot (\theta ) \partial _{\theta \phi } T_{{rr}}{}^{{(0)}} +\csc ^2(\theta ) \partial _{\phi \phi \phi } T_{{rr}}{}^{{(0)}} +\partial _{\theta \theta \phi } T_{{rr}}{}^{{(0)}}-6 \partial _{\phi } T_{{rr}}{}^{{(0)}} -\right .\\
& 8 \partial _{\phi } T_{{ur}}{}^{{(0)}} -16 \cot (\theta ) \partial _{\phi } T_{{r\theta }}{}^{{(1)}} -4 \partial _{\theta \phi } T_{{r\theta }}{}^{{(1)}} -10 \csc ^2(\theta ) \partial _{\phi \phi } T_{{r\phi }}{}^{{(1)}} +\\
& \left .6 \cot (\theta ) \partial _{\theta } T_{{r\phi }}{}^{{(1)}} -6 \partial _{\theta \theta } T_{{r\phi }}{}^{{(1)}} -36 T_{{r\phi }}{}^{{(1)}}\right)+\\
& \frac{1}{32 \pi }\left(3 \cot (\theta ) \partial _{\theta \phi } \gamma ^{{(2)}} +(5 \cos (2 \theta )-1) \csc ^2(\theta ) \partial _{\phi } \gamma ^{{(2)}} -\csc ^2(\theta ) \partial _{\phi \phi \phi } \gamma ^{{(2)}} -\right .\\
& \partial _{\theta \theta \phi } \gamma ^{{(2)}}+3 \cos (\theta ) \partial _{\theta \theta } \delta ^{{(2)}} +\sin (\theta ) \partial _{\theta \theta \theta } \delta ^{{(2)}} +\csc (\theta ) \partial _{\theta \phi \phi } \delta ^{{(2)}} +\\
& \left .4 \cot (\theta ) \csc (\theta ) \partial _{\phi \phi } \delta ^{{(2)}} +12 \delta ^{{(2)}} \cos (\theta )+(-2 \cos (2 \theta )-1) \csc (\theta ) \partial _{\theta } \delta ^{{(2)}} \right),
\end{aligned}
\end{equation}
\begin{equation}\label{Tuu0}
\begin{aligned}
T_{uu}{}^{(0)}=\;&
\frac{1}{48} \left(\left(\cot ^2(\theta )+8\right) \partial _{\theta \theta } T_{{rr}}{}^{{(0)}} -2 \cot (\theta ) \partial _{\theta \theta \theta } T_{{rr}}{}^{{(0)}} -2 \csc ^2(\theta ) \partial _{\theta \theta \phi \phi } T_{{rr}}{}^{{(0)}} +\right .\\
& 2 \cot (\theta ) \csc ^2(\theta ) \partial _{\theta \phi \phi } T_{{rr}}{}^{{(0)}} -4 \cos (2 \theta ) \csc ^4(\theta ) \partial _{\phi \phi } T_{{rr}}{}^{{(0)}} -\\
& \csc ^4(\theta ) \partial _{\phi \phi \phi \phi } T_{{rr}}{}^{{(0)}} -(3 \cos (2 \theta )-2) \cot (\theta ) \csc ^2(\theta ) \partial _{\theta } T_{{rr}}{}^{{(0)}} -\partial _{\theta \theta \theta \theta } T_{{rr}}{}^{{(0)}}+\\
& 8 \csc ^2(\theta ) \partial _{\phi \phi } T_{{ur}}{}^{{(0)}} +8 \cot (\theta ) \partial _{\theta } T_{{ur}}{}^{{(0)}} +8 \partial _{\theta \theta } T_{{ur}}{}^{{(0)}} -48 T_{{ur}}{}^{{(0)}}+\\
& 20 \cot (\theta ) \partial _{\theta \theta } T_{{r\theta }}{}^{{(1)}} +10 \csc ^2(\theta ) \partial _{\theta \phi \phi } T_{{r\theta }}{}^{{(1)}} +10 \cot (\theta ) \csc ^2(\theta ) \partial _{\phi \phi } T_{{r\theta }}{}^{{(1)}} -\\
& 2 T_{{r\theta }}{}^{{(1)}} (9 \cos (2 \theta )-14) \cot (\theta ) \csc ^2(\theta )-(13 \cos (2 \theta )-3) \csc ^2(\theta ) \partial _{\theta } T_{{r\theta }}{}^{{(1)}} +\\
& 10 \partial _{\theta \theta \theta } T_{{r\theta }}{}^{{(1)}} +10 \csc ^2(\theta ) \partial _{\theta \theta \phi } T_{{r\phi }}{}^{{(1)}} -30 \cot (\theta ) \csc ^2(\theta ) \partial _{\theta \phi } T_{{r\phi }}{}^{{(1)}} -\\
&\left . 8 (\cos (2 \theta )-6) \csc ^4(\theta ) \partial _{\phi } T_{{r\phi }}{}^{{(1)}} +10 \csc ^4(\theta ) \partial _{\phi \phi \phi } T_{{r\phi }}{}^{{(1)}} \right)+\\
& \frac{1}{128 \pi }\left((\cos (2 \theta )+5) \csc ^2(\theta ) \partial _{\theta \theta } \gamma ^{{(2)}} -8 \cot (\theta ) \partial _{\theta \theta \theta } \gamma ^{{(2)}} -\right .\\
& 12 \cot (\theta ) \csc ^2(\theta ) \partial _{\theta \phi \phi } \gamma ^{{(2)}} +2 (7-3 \cos (2 \theta )) \csc ^4(\theta ) \partial _{\phi \phi } \gamma ^{{(2)}} +\\
& 2 \csc ^4(\theta ) \partial _{\phi \phi \phi \phi } \gamma ^{{(2)}} +2 (8 \cos (2 \theta )-11) \cot (\theta ) \csc ^2(\theta ) \partial _{\theta } \gamma ^{{(2)}} -2 \partial _{\theta \theta \theta \theta } \gamma ^{{(2)}} +\\
& 24 \gamma ^{{(2)}}-4 \csc (\theta ) \partial _{\theta \theta \theta \phi } \delta ^{{(2)}} +2 (7 \cos (2 \theta )-3) \csc ^3(\theta ) \partial _{\theta \phi } \delta ^{{(2)}} -\\
& 4 \csc ^3(\theta ) \partial _{\theta \phi \phi \phi } \delta ^{{(2)}} +4 (3 \cos (2 \theta )-7) \cot (\theta ) \csc ^3(\theta ) \partial _{\phi } \delta ^{{(2)}} -\\
&\left . 4 \cot (\theta ) \csc ^3(\theta ) \partial _{\phi \phi \phi } \delta ^{{(2)}} \right).
\end{aligned}
\end{equation}

\end{appendices}

\bibliography{refs}

\end{document}